\documentclass[fleqn,usenatbib]{mnras}

\usepackage[T1]{fontenc}

\DeclareRobustCommand{\VAN}[3]{#2}
\let\VANthebibliography\thebibliography
\def\thebibliography{\DeclareRobustCommand{\VAN}[3]{##3}\VANthebibliography}

\usepackage{graphicx}	
\usepackage{amsmath}	
\usepackage{amssymb}	
\usepackage{xcolor}
\usepackage{txfonts}

\newcommand\T{\rule{0pt}{2.6ex}}       
\newcommand\B{\rule[-1.2ex]{0pt}{0pt}} 

\newcommand{\beq}{\begin{equation}}
\newcommand{\eeq}{\end{equation}}

\def\gs{\mathrel{\lower0.6ex\hbox{$\buildrel {\textstyle >}\over{\scriptstyle \sim}$}}}
\def\ls{\mathrel{\lower0.6ex\hbox{$\buildrel {\textstyle <}\over{\scriptstyle \sim}$}}}
\newcommand{\simgt}{\lower.5ex\hbox{$\; \buildrel > \over \sim \;$}}
\newcommand{\simlt}{\lower.5ex\hbox{$\; \buildrel < \over \sim \;$}}

\defcitealias{se+et15_comalit_I}{CoMaLit-I}
\defcitealias{ser+al15_comalit_II}{CoMaLit-II}
\defcitealias{ser15_comalit_III}{CoMaLit-III}
\defcitealias{se+et15_comalit_IV}{CoMaLit-IV}
\defcitealias{se+et17_comalit_V}{CoMaLit-V}
\defcitealias{xxl_I_pie+al16}{XXL~Paper~I}
\defcitealias{xxl_II_pac+al16}{XXL~Paper~II}
\defcitealias{xxl_III_gil+al16}{XXL~Paper~III}
\defcitealias{xxl_IV_lie+al16}{XXL~Paper~IV}
\defcitealias{xxl_XIII_eck+al16}{XXL~Paper~XIII}
\defcitealias{xxl_XX_ada+al18}{XXL~Paper~XX}
\defcitealias{xxl_XXV_pac+al18}{XXL~Paper~XXV}
\defcitealias{xxl_XXXVIII_ser+al19}{XXL~Paper~XXXVIII}

   \defcitealias{pierre2016new}{XXL~Paper~I}
   \defcitealias{pacaud2016new}{XXL~Paper~II}
   \defcitealias{2016A&A...592A...3G}{XXL~Paper~III}
   \defcitealias{2016A&A...592A...4L}{XXL~Paper~IV}
   \defcitealias{2014ApJ...794..157M}{XXL~Paper~V}
   \defcitealias{2016A&A...592A...5F}{XXL~Paper~VI}
   \defcitealias{2016A&A...592A...6P}{XXL~Paper~VII}
   \defcitealias{2016A&A...592A...7A}{XXL~Paper~VIII}
   \defcitealias{2016A&A...592A...8B}{XXL~Paper~IX}
   \defcitealias{2016A&A...592A...9Z}{XXL~Paper~X}
   \defcitealias{2016A&A...592A..10S}{XXL~Paper~XI}
   \defcitealias{2016A&A...592A..11K}{XXL~Paper~XII}
   \defcitealias{eckert2016}{XXL~Paper~XIII}
   \defcitealias{2016PASA...33....1L}{XXL~Paper~XIV}
   \defcitealias{2016MNRAS.462.4141L}{XXL~Paper~XV}
   \defcitealias{2018A&A...620A...1M}{XXL~Paper~XVI}
   \defcitealias{2018A&A...620A...2M}{XXL~Paper~XVII}
   \defcitealias{2018A&A...620A...3B}{XXL~Paper~XVIII}
   \defcitealias{koulouridis2018b}{XXL~Paper~XIX}
   \defcitealias{adami2018}{XXL~Paper~XX}
   \defcitealias{2018A&A...620A...6M}{XXL~Paper~XXI}
   \defcitealias{2018A&A...620A...7G}{XXL~Paper~XXII}
   \defcitealias{2018A&A...620A...8F}{XXL~Paper~XXIII}
   \defcitealias{2018A&A...620A...9F}{XXL~Paper~XXIV}
   \defcitealias{2018A&A...620A..10P}{XXL~Paper~XXV}
   \defcitealias{2018A&A...620A..11C}{XXL~Paper~XXVI}
   \defcitealias{Chiappetti2018}{XXL~Paper~XXVII}
   \defcitealias{2018A&A...620A..13R}{XXL~Paper~XXVIII}
   \defcitealias{2018A&A...620A..14S}{XXL~Paper~XXIX}
   \defcitealias{2018A&A...620A..15G}{XXL~Paper~XXX}
   \defcitealias{2018A&A...620A..16B}{XXL~Paper~XXXI}
   \defcitealias{2018A&A...620A..17P}{XXL~Paper~XXXII}
   \defcitealias{2018A&A...620A..18L}{XXL~Paper~XXXIII}
   \defcitealias{faccioli2018}{XXL~Paper~XXIV}
   \defcitealias{koulouridis2018}{XXL~Paper~XXXV}

\title[Understanding the selection of galaxy clusters]{Understanding X-ray and optical selection of galaxy clusters: A comparison of the XXL and CAMIRA cluster catalogues obtained in the common XXL-HSC SSP area}

\author[J. P. Willis et al.]{
J. P. Willis,$^{1}$\thanks{E-mail: jwillis@uvic.ca (JPW)}
M. Oguri,$^{2,3,4}$
M. E. Ramos-Ceja,$^{5,6}$
F. Gastaldello,$^{7}$
M. Sereno,$^{8,9}$
C. Adami,$^{10}$
S. Alis,$^{11}$\newauthor
B. Altieri,$^{12}$
L. Chiappetti,$^{13}$
P.S. Corasaniti$^{14,15}$
D. Eckert,$^{16}$
S. Ettori,$^{8}$
C. Garrel,$^{17,18}$
P. Giles,$^{19}$
J. Lefevre$^{17}$\newauthor
L. Faccioli,$^{18}$
S. Fotopoulou,$^{20}$
A. Hamabata,$^{3}$
E. Koulouridis,$^{21}$
M. Lieu,$^{12}$
Y.-T. Lin,$^{22}$
B. Maughan,$^{20}$\newauthor
A. J. Nishizawa,$^{23,24}$
T. Okabe,$^{3}$
N. Okabe,$^{25,26,27}$
F. Pacaud,$^{5}$
S. Paltani,$^{16}$
M. Pierre,$^{18}$
M. Plionis,$^{21,28}$\newauthor
B. Poggianti,$^{29}$
E. Pompei,$^{30}$
T. Sadibekova$^{15}$
K. Umetsu,$^{22}$
and P. Valageas,$^{31}$
\\
$^{1}$Department of Physics and Astronomy, University of Victoria, 3800 Finnerty Road, Victoria, V8P 5C2, BC, Canada\\
$^{2}$Research Center for the Early Universe, The University of Tokyo, 7-3-1 Hongo, Bunkyo-ku, Tokyo 113-0033, Japan\\ 
$^{3}$Department of Physics, The University of Tokyo, 7-3-1 Hongo, Bunkyo-ku, Tokyo 113-0033, Japan\\
$^{4}$Kavli Institute for the Physics and Mathematics of the Universe (Kavli IPMU, WPI), The University of Tokyo, 5-1-5 Kashiwanoha, Kashiwa, Chiba 277-8582, Japan\\
$^{5}$Argelander-Institut f{\" u}r Astronomie, University of Bonn, Auf dem H{\" u}gel 71, 53121 Bonn, Germany\\
$^{6}$Max-Planck Institut f\"ur extraterrestrische Physik, Postfach 1312, 85741 Garching bei M\"unchen, Germany\\
$^{7}$INAF - IASF Milano, via A. Corti 12, 20133 Milano, Italy\\
$^{8}$INAF -- Osservatorio di Astrofisica e Scienza dello Spazio di Bologna, via Piero Gobetti 93/3, I-40129 Bologna, Italy\\
$^{9}$INFN, Sezione di Bologna, viale Berti Pichat 6/2, 40127 Bologna, Italy\\
$^{10}$Aix Marseille Univ, CNRS, CNES, LAM, Marseille, France\\
$^{11}$Department of Astronomy and Space Sciences, Faculty of Science, Istanbul University, 34119 Istanbul, Turkey\\
$^{12}$European Space Astronomy Centre, ESA, Villanueva de la Ca{\~ n}ada, 28691 Madrid, Spain\\
$^{13}$INAF, IASF Milano, via Bassini 15, Milano, I-20133, Italy\\
$^{14}$LUTH, UMR 8102 CNRS, Observatoire de Paris, PSL Research University, Universit{\' e} Paris Diderot, 5 Place Jules Janssen, F-92190 Meudon, France\\
$^{15}$Sorbonne Universit{\' e}, CNRS, UMR 7095, Institut d'Astrophysique de Paris, 98 bis bd Arago, F-75014 Paris, France\\
$^{16}$Department of Astronomy, University of Geneva, 1205, Versoix, Switzerland\\
$^{17}$IRFU, CEA, Universit{\' e} Paris-Saclay, 91191, Gif-sur-Yvette, France\\ 
$^{18}$Universit{\' e} Paris Diderot, AIM, Sorbonne Paris Cit{\' e}, CEA, CNRS, 91191, Gif-sur-Yvette, France\\
$^{19}$Department of Physics and Astronomy, University of Sussex, Falmer, Brighton BN1 9QH, UK\\
$^{20}$HH Wills Physics Laboratory, University of Bristol, Tyndall Ave, Bristol, BS8 1TL, UK\\
$^{21}$Institute for Astronomy \& Astrophysics, Space Applications \& Remote Sensing, National Observatory of Athens, GR-15236 Palaia Penteli, Greece\\
$^{22}$Academia Sinica Institute of Astronomy and Astrophysics (ASIAA), No. 1,
Section 4, Roosevelt Road, Taipei 10617, Taiwan\\
$^{23}$Institute for Advanced Research, Nagoya University, Furocho, Chikusa-ku, Nagoya, Aichi, 464-8602, Japan\\
$^{24}$Department of Science, Nagoya University, Furocho, Chikusa-ku, Nagoya, Aichi, 464-8602, Japan\\
$^{25}$Physics Program, Graduate School of Advanced Science and Engineering, Hiroshima University, 1-3-1 Kagamiyama, Higashi-Hiroshima, Hiroshima 739-8526, Japan\\
$^{26}$Hiroshima Astrophysical Science Center, Hiroshima University, 1-3-1 Kagamiyama, Higashi-Hiroshima, Hiroshima 739-8526, Japan\\
$^{27}$Core Research for Energetic Universe, Hiroshima University, 1-3-1, Kagamiyama, Higashi-Hiroshima, Hiroshima 739-8526, Japan\\
$^{28}$Physics Department, Aristotle University of Thessaloniki, Thessaloniki, 54124, Greece\\
$^{29}$INAF-Padova Astronomical Observatory, Vicolo dell'Osservatorio 5, I-35122 Padova, Italy\\
$^{30}$ESO-Chile, Alonso de Cordova 3107, Vitacura, Chile\\
$^{31}$Institut de Physique Th{\' e}orique, Universit{\' e} Paris-Saclay, CEA, CNRS, F-91191 Gif-sur-Yvette Cedex, France
}

\date{Accepted XXX. Received YYY; in original form ZZZ}

\pubyear{2015}

\begin{document}
\label{firstpage}
\pagerange{\pageref{firstpage}--\pageref{lastpage}}
\maketitle

\begin{abstract}
Large samples of galaxy clusters provide knowledge of both
astrophysics in the most massive virialised environments and the
properties of the cosmological model that defines our
Universe. However, an important issue that affects the interpretation
of galaxy cluster samples is the role played by the selection waveband
and the potential for this to introduce a bias in the physical
properties of clusters thus selected.  We aim to investigate
waveband-dependent selection effects in the identification of galaxy
clusters by comparing the X-ray Multi-Mirror (XMM) Ultimate
Extra-galactic Survey (XXL) and Subaru Hyper Suprime-Cam (HSC) CAMIRA
cluster samples identified from a common 22.6 deg$^2$ sky area.  We
compare 150 XXL and 270 CAMIRA clusters in a common parameter space
defined by X-ray aperture brightness and optical richness. We find
that 71/150 XXL clusters are matched to the location of a CAMIRA
cluster, the majority of which (67/71) display richness values $N>15$
that exceed the CAMIRA catalogue richness threshold.  We find that
67/270 CAMIRA clusters are matched to the location of an XXL cluster
(defined within XXL as an extended X-ray source). Of the unmatched
CAMIRA clusters, the majority display low X-ray fluxes consistent with
the lack of an XXL counterpart.  However, a significant fraction
(64/107) CAMIRA clusters that display high X-ray fluxes are not
asociated with an extended source in the XXL catalogue.  We
demonstrate that this disparity arises from a variety of effects
including the morphological criteria employed to identify X-ray
clusters and the properties of the XMM PSF.
\end{abstract}




\section{Introduction}

The identification of large samples of galaxy clusters from
observations compiled at various wavelengths represents a mature field
of study. Wavebands and techniques employed to identify galaxy
clusters include the identification of spatial overdensities of
galaxies displaying characteristically-red colours from optical and
NIR imaging data
\citep{postman1996,gladders2000,rykoff2014,oguri2014,maturi2019}, the
detection of optically-thin X-ray photons resulting from
bremsstrahlung emission from the hot, baryonic intra-cluster medium
(ICM; \citealt{gioia1990,bohringer2001,clerc2012}), the observation of
a Sunyaev Zel'dovich (SZ) decrement caused by inverse Compton
scattering of cosmic microwave background (CMB) photons by electrons
in the cluster ICM
\citep{stan2009,marriage2011,reichardt2013,planck2016}, and the
detection of weak lensing shear in the images of background galaxies
arising from the cluster gravitational potential
\citep{miyazaki2002,wittman2006,gavazzi2007,miyazaki2018b}.
 
One aspect of the study of galaxy clusters that is less well
understood however, is the relationship between the observing waveband
and the average physical properties of cluster samples thus generated.
Much work has been undertaken to understand the multi-wavelength
properties of galaxy clusters detected in a given waveband and, in
particular, to express these properties via scaling relationships
\citep[e.g.][]{rozo2014a,rozo2014b}. An associated approach attempts to
understand the combination of effects that lead to a given fraction of
galaxy clusters within a sample being detected in one waveband but not
another, e.g. an optically identified galaxy cluster not being
detected in X-ray \citep{donahue2002,sadibekova2014}. Further
understanding is achieved by performing detailed multi-wavelength
follow-up studies of galaxy clusters identified in a particular
waveband \citep[e.g.][]{rossetti2017,santos2017,zhang2019}.

We have previously considered aspects of this question in
\citet{willis2018} where we compared the physical properties of two
distant cluster samples: the X-ray selected XMM-LSS survey and the
optical-MIR selected SpARCS sample. The results of this comparison
indicated that many of the observed differences between the two
cluster samples could be interpreted in terms of a larger uncertainty
in the centroid estimation resulting from MIR galaxy overdensity
compared to X-ray emission. Furthermore, MIR selected clusters were
found to have marginally more numerous red sequence populations
compared to X-ray selected clusters of comparable X-ray
brightness. Ultimately, the relatively small number of clusters
compared (19 XMM-LSS and 92 SpARCS) limited the extent to which
physical differences between the two samples could be resolved. This
led us to seek a more comprehensive comparison, between cluster
samples from the XXL X-ray survey \citep[][hereafter
  \citetalias{pierre2016new}]{pierre2016new} and the Subaru Hyper
Suprime-Cam (HSC) optical imaging survey known as HSC Subaru Strategic
Program \citep[HSC-SSP;][]{miyazaki2018a,aihara2018a}, as presented in
this paper.

The structure of this paper is as follows: In Section
\ref{sec_samples} we describe the two cluster samples and perform a
simple matching analysis. We then compute scaling relations for each
sample prior to defining cluster sub-samples on the basis of X-ray
aperture photometry and cluster richness measurements. In Section
\ref{sec_results} we compile a number of physical measurements for
each cluster sub-sample before moving to Section \ref{sec_discussion}
where we discuss and attempt to explain the nature of the physical
differences between each cluster sub-sample. We draw our conclusions
in Section \ref{sec_conc}. Where necessary, we assume a
Friedmann-Lema{\^ i}tre-Robertson-Walker cosmological model described
by the parameters $\Omega_M = 0.3$, $\Omega_\Lambda = 0.7$, $H_0=70 \,
\rm kms^{-1}Mpc^{-1}$. In this model a transverse physical scale of 700 kpc
observed at a redshift $z=0.6$ corresponds to an angular scale of 1.75
arcminutes.

\section{The cluster samples}
\label{sec_samples}

The XXL sample employed in this paper consists of 150 clusters
presented within \citealt{adami2018} (hereafter
\citetalias{adami2018}).  This version of the XXL catalogue results
from the processing of individual XMM pointings with version 3.3 of
the {\tt Xamin} pipeline and is limited to sources at XMM off axis
angles $<13$ arcminutes.  Clusters in this catalogue have $0.1 < z <
1.3$ and are selected as either class 1 or class 2 (C1 and C2)
extended sources on the basis of their surface brightness
characteristics as defined by \citet{pacaud2006} and
\citet{pacaud2016new} (hereafter \citetalias{pacaud2016new}). Sources
for which a point source model produces a statistically acceptable fit
are labelled as P1.  The remaining sources for which neither an
extended source model (C1 or C2) nor a point source model (P1) produce
an acceptable characterisation are labelled as P0 in the XXL database.
Though such sources typically lack sufficient X-ray photon counts to
generate a statistically acceptable fit the full sample of P0 and P1
sources expected to be dominated numerically by extra-galactic X-ray
point sources.  In the following discussion, we refer to both P0 and
P1 sources from the XXL version 3.3 catalogue as point sources,
although we recognise that individual point sources may represent
faint extended sources where low source counts prevent a statistically
acceptable classification. This issue is potentially of importance for
the case of X-ray point sources studied along the line of sight to
clusters detected in optical wavebands. In such cases the cluster
detection effectively acts as a prior selection applied to the P0 and
P1 sample.

Of the 150 XXL clusters, 142 are confirmed spectroscopically whereas
the remaining 8 clusters possess a photometric redshift
(\citetalias{adami2018}). The number of XXL clusters employed in this
paper is slightly greater than that used in the joint HSC-XXL weak
lensing study of \citet[150 compared to 136 clusters]{umetsu2020} as
in the present paper we select clusters from a common sky area with no
prior selection based upon relative cluster positions.

The HSC sample consists of 289 clusters around the XXL region selected
from the S17A data release \citep{aihara2018b,aihara2019} employing
the CAMIRA red-sequence detection algorithm
\citep{oguri2014,oguri2018}. Of these, 270 clusters lie within 13
arcminutes of an XMM pointing.  Detected clusters are characterised by
a red-sequence derived photometric redshift and a stellar-mass
corrected richness ($N$) measured using a spatially extended filter of
radial scale 0.8 Mpc \citep[see][and Section \ref{sec_rich} for more
  details]{oguri2014}.  The photometric redshift accuracy of the
CAMIRA catalogue is estimated to be $\Delta z/(1+z) < 0.01$
\citep{oguri2018}.  The CAMIRA sample is restricted to $0.1 < z <
1.38$ and richness $N>15$.

The common sky area between the XXL and HSC-SSP surveys was
computed using a method similar to that presented in \citep[][their
  Figure 1]{umetsu2020}, i.e. we computed the overlap between the
HSC-SSP survey and the grid of XXL XMM pointing centres with the
additional constraint that only the area with 13 arcminutes of each
XMM pointing centre contributed to the area calculation. Using this
method we obtain a common sky area of 22.6 square degrees.

\subsection{X-ray aperture photometry}
\label{sec_apphot}

The common sky area of each cluster sample has been observed by
XMM-Newton as part of the XXL survey which consists of a contiguous
field of 10 ks XMM exposures \citepalias{pierre2016new}. We performed
X-ray aperture photometry in the [0.5-2] keV waveband at the sky
location of all clusters following the procedure described in
\citet{willis2018}, i.e. apertures are placed at the X-ray centroid
for XXL clusters and at the optical centroid for CAMIRA clusters.
X-ray photometry was performed in a circular aperture of radius 500
kpc which corresponds to a scale approximately $0.9 \times
r_{500}$\footnote{Where $r_{500}$ is defined as the physical radius
  within which the average cluster density exceeds 500 times the
  critical density of the universe at that redshift.}  inferred for
X-ray bright XXL clusters \citep{umetsu2020}.  In particular, point sources
were excluded from the aperture photometry with a purely geometric
correction applied to account for the reduced area sampling.
Point source locations were obtained from the {\tt Xamin} pipeline and
represent all X-ray sources not classified as C1 or C2
\citep[][hereafter \citetalias{faccioli2018}]{faccioli2018}.  The {\tt
  Xamin} pipeline employs {\tt SExtractor} \citep{bertin1996} to
create a segmentation map that is used to mask each point source,
whose extent varies but is normally much smaller than the 500 kpc
rest-frame aperture size used for photometry. We further
compute the X-ray luminosity of each cluster ($L_X$) employing the
aperture flux, a distance modulus calculated from the cluster redshift
and a $k$-correction based upon a standard, $T=2$ keV plasma emission
model \citep{willis2018}.

\begin{figure}
\centering
\includegraphics[width=3.5in]{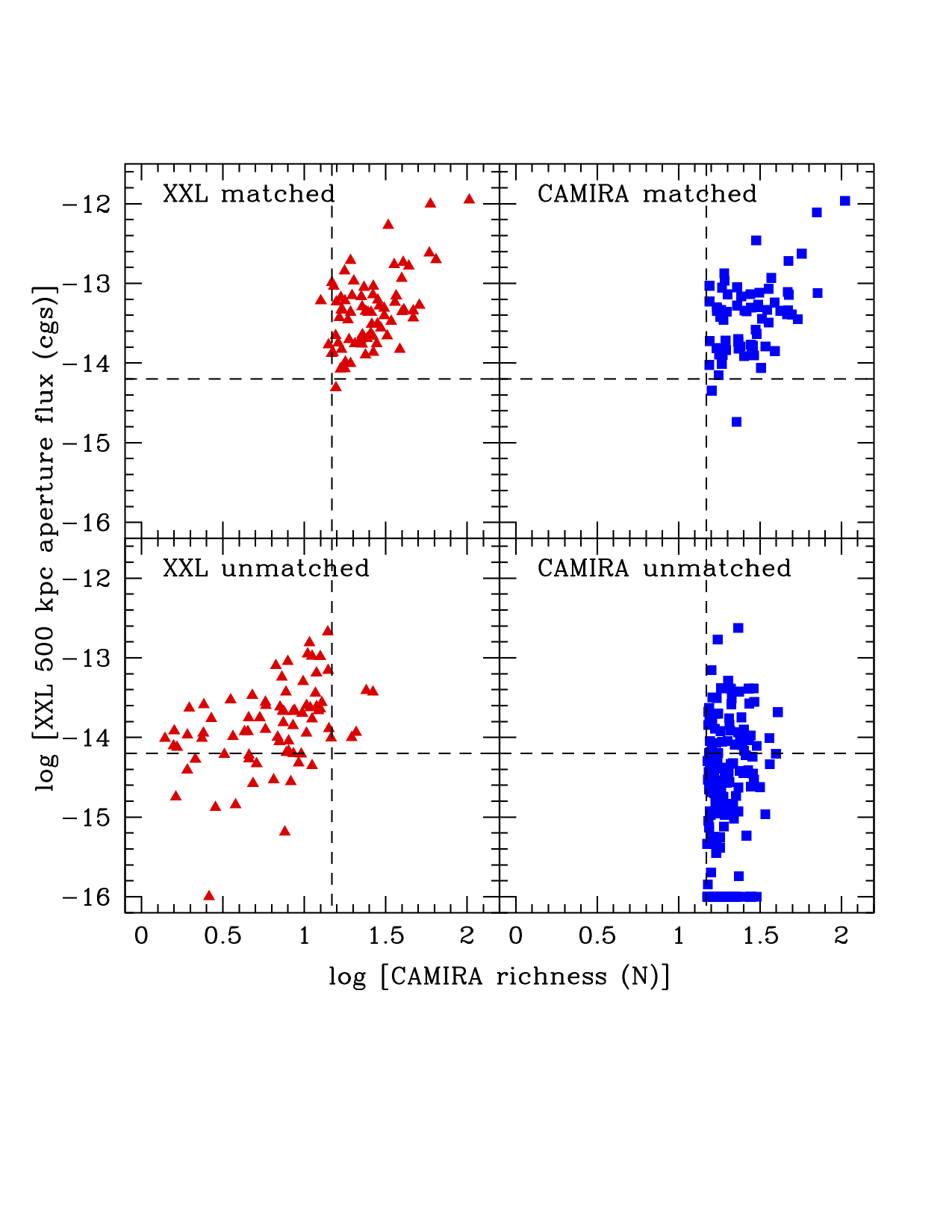}
\caption{X-ray aperture flux versus richness for XXL (red) and CAMIRA
  (blue) clusters. The vertical dashed line in each panel indicates
  the $N=15$ richness cut applied to generate the CAMIRA sample while
  the horizontal dashed line indicates $\log f_X=-14.2$ (cgs), an
  assigned threshold to approximately mimic the distribution of $N>15$
  XXL clusters. The horizontal group of points at $\log f_X=-16$ (cgs)
  represents X-ray undetected sources.}
\label{fig_flux_rich_all}
\end{figure}

\subsection{Computing CAMIRA richness values for XXL clusters}
\label{sec_rich}

We follow the standard algorithm in CAMIRA to compute richness (Oguri
2014). For each peak in a three-dimentional richness map, it first
identifies a central cluster galaxy (CCG) candidate that maximizes
the likelihood function consisting of the stellar mass filter, the
member galaxy likelihood, and spatial filter, such that a massive
galaxy located in the red-sequence and within $\la 0.3h^{-1}$Mpc from
the peak is selected as a CCG candidate. After the CCG candidate is
identified, CAMIRA re-computes the cluster photometric redshift by
combining photometric redshift estimates of red-sequence galaxies
around the CCG candidate and re-computes richness by summing up the
number parameter of galaxies around the CCG candidate with a spatial
filter of $F_R(R) \propto \Gamma[n/2,
  (R/R_0)^2]-(R/R_0)^ne^{-(R/R_0)^2}$ with $n=4$, $R_0=0.8h^{-1}$Mpc,
and $\Gamma$ being Gamma function (see Oguri 2014 for more
details). With this spatial filter, the number of galaxies within $\la
1h^{-1}$Mpc is used to define the richness. The spatial filter is a
compensated filter and thus subtracts the background level from the
number density of red galaxies around each cluster. We again search
for a new CCG candidate with the new center and the cluster redshift,
and repeat the process mentioned above until it converges.

As in the case of X-ray aperture photometry, here we want to compute
richness for all the XXL clusters. To do so, we simply replace peaks
in the three-dimentional richness map with X-ray centroids and
redshifts of XXL clusters and compute the richness for each XXL
cluster using the same procedure as mentioned above.

\subsection{Matching results and the definition of cluster subsamples}
\label{sec_subsamp}

Figure \ref{fig_flux_rich_all} shows 500 kpc aperture X-ray flux
($f_X$) versus CAMIRA measured richness ($N$) for all clusters.  To
further our understanding of the XXL and CAMIRA samples clusters from
each catalogue were matched according to a positional and redshift
tolerance.  Cluster detections were considered to be matched if they
displayed a rest-frame transverse physical offset within 700 kpc
(computed at the redshift of the cluster about which a match was being
sought).  In addition we applied the criterion that the difference
between the XXL and CAMIRA catalogue redshifts should be $\Delta z <
0.1$.  Matching results are summarised in Table \ref{tab_match} and
employ cluster sub-samples defined in the following discussion.
\begin{table}
\caption{Matching results between the XXL (150 objects) and CAMIRA (270 objects) cluster samples}
\label{tab_match}      
\centering                          
\begin{tabular}{|l|c|c|}        
\hline\hline                 
Sample & matched & unmatched \\
\hline
XXL $N>15$ & 67/71 & 4/71 \\
XXL $N<15$ & 0/79 & 79/79 \\
\hline
CAMIRA $\log f_X>-14.2$ (cgs) & 64/107 & 43/107 \\
CAMIRA $\log f_X<-14.2$ (cgs) & 3/163 & 160/163 \\
\hline                                   
\end{tabular}
\end{table}

It is immediately apparent that approximately all XXL clusters that
display a richness of $N>15$ are matched to a cluster in the
CAMIRA catalogue. However, the converse is not true, a sizeable
fraction of CAMIRA clusters (displaying $N>15$ by definition) that are
of comparable X-ray aperture flux to XXL clusters are not matched to a
XXL cluster. Note that, as we discuss in Section \ref{sec_apphot},
despite not being matched to an XXL C1 or C2 cluster, many unmatched
CAMIRA clusters have detectable X-ray emission. Determining the
physical cause of this apparent disparity motivates the remainder of
the paper.

Although the XXL cluster sample is limited by X-ray surface brightness
(\citealt{pacaud2006}; \citetalias{pacaud2016new}), it can reasonably
be approximated to a flux limited sample at fixed core radius
\citep[Figure 8 of][]{pacaud2006}. We therefore apply a flux limit of
$\log f_X = -14.2$ (cgs) to the CAMIRA cluster sample (see Figure
\ref{fig_flux_rich_all}).  Note that this limit is approximately two
times fainter than the value of $\log f_X = -13.8$ (cgs) corresponding
to the 100\% XMM on-axis completeness limit presented by
\citetalias{adami2018}\footnote{Note that \citetalias{adami2018}
measure fluxes within a 1 arcminute radius circular aperture compared
to the 500 kpc radius aperture (1.25 arcminutes at $z=0.6$ using the
adopted cosmological model) employed in this paper.}. The limit of
$\log f_X = -14.2$ (cgs) presented in this paper selects 107 CAMIRA
clusters which we refer to as ``high flux'' in the following
discussion. Of these high flux CAMIRA clusters, 64 are matched to an
XXL cluster (see Table \ref{tab_match}).

\begin{figure}
\centering
\includegraphics[width=3.5in]{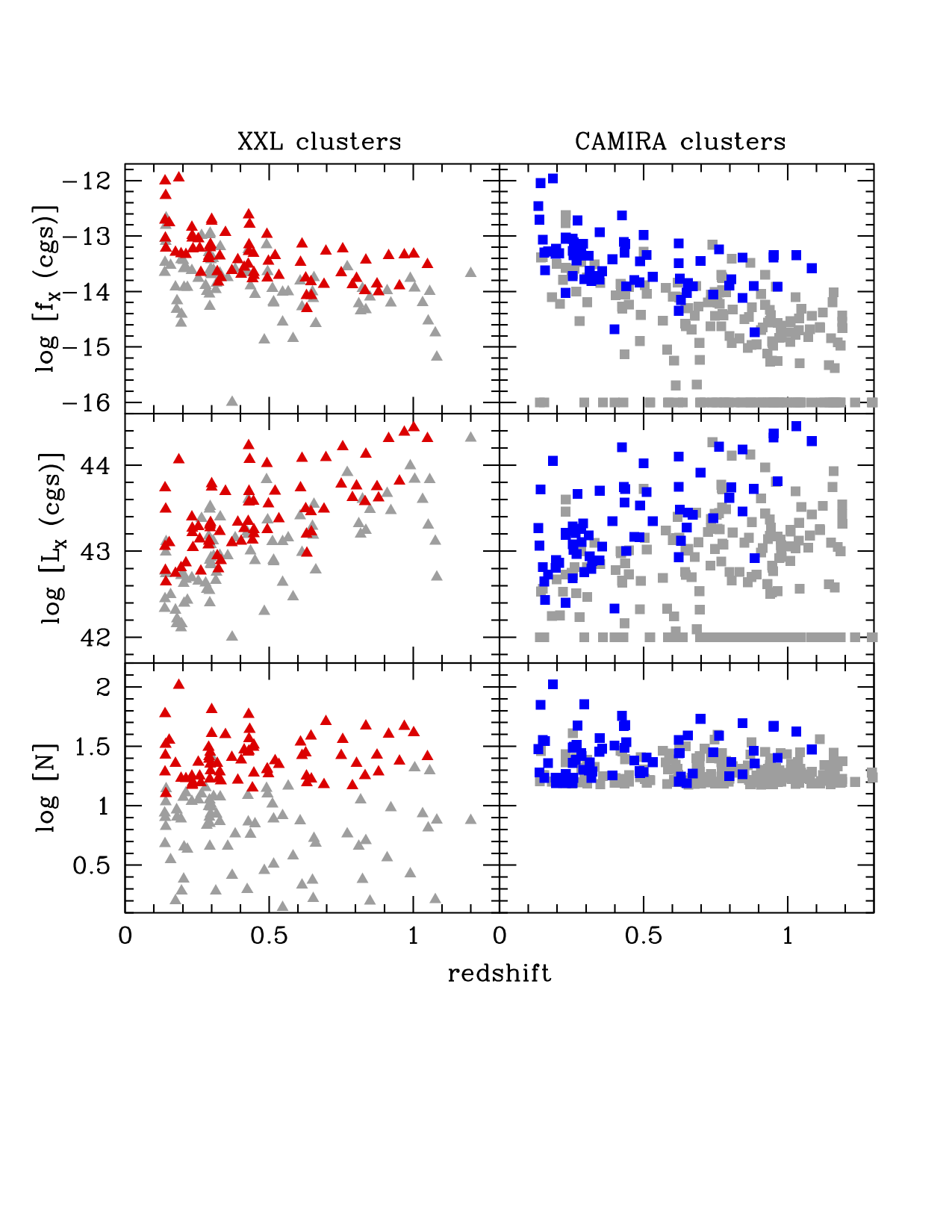}
\caption{The distribution of X-ray flux (top panels), luminosity
  (centre panels) and CAMIRA richness (bottom panels) versus redshift
  for XXL (left panels; red symbols) and CAMIRA (right panels; blue
  symbols) clusters. Red/blue symbols in all panels denote matched
  clusters and grey symbols denote unmatched clusters. The horizontal
  group of points at $\log f_X=-16$ (cgs) in the top right panel and
  $\log L_X=42$ (cgs) in the centre right panel represents X-ray
  undetected sources.}
\label{fig_fx_lx_n_red}
\end{figure}

Of the 163 clusters that lie below this flux limit, which we refer to
as ``low flux'' in the following discussion, only 3 are matched to an
XXL cluster. The simplest explanation is that these low flux clusters
are too faint to be unambiguously flagged as extended sources by the
XXL pipeline. Though such sources may represent true extended sources,
there are insufficient X-ray photons to permit a statistically
acceptable characterisation. Such sources are labelled P0 in the XXL
catalogue. Figure \ref{fig_fx_lx_n_red} displays the trends of flux,
luminosity and richness in both the XXL and CAMIRA samples and
demonstrates that the low flux CAMIRA clusters are high to moderate
luminosity clusters viewed at high ($z>0.5$) redshift.  Of the 107
high-flux CAMIRA clusters, 43 remain unmatched to an XXL cluster.

\begin{figure}
\centering
\includegraphics[width=3.5in]{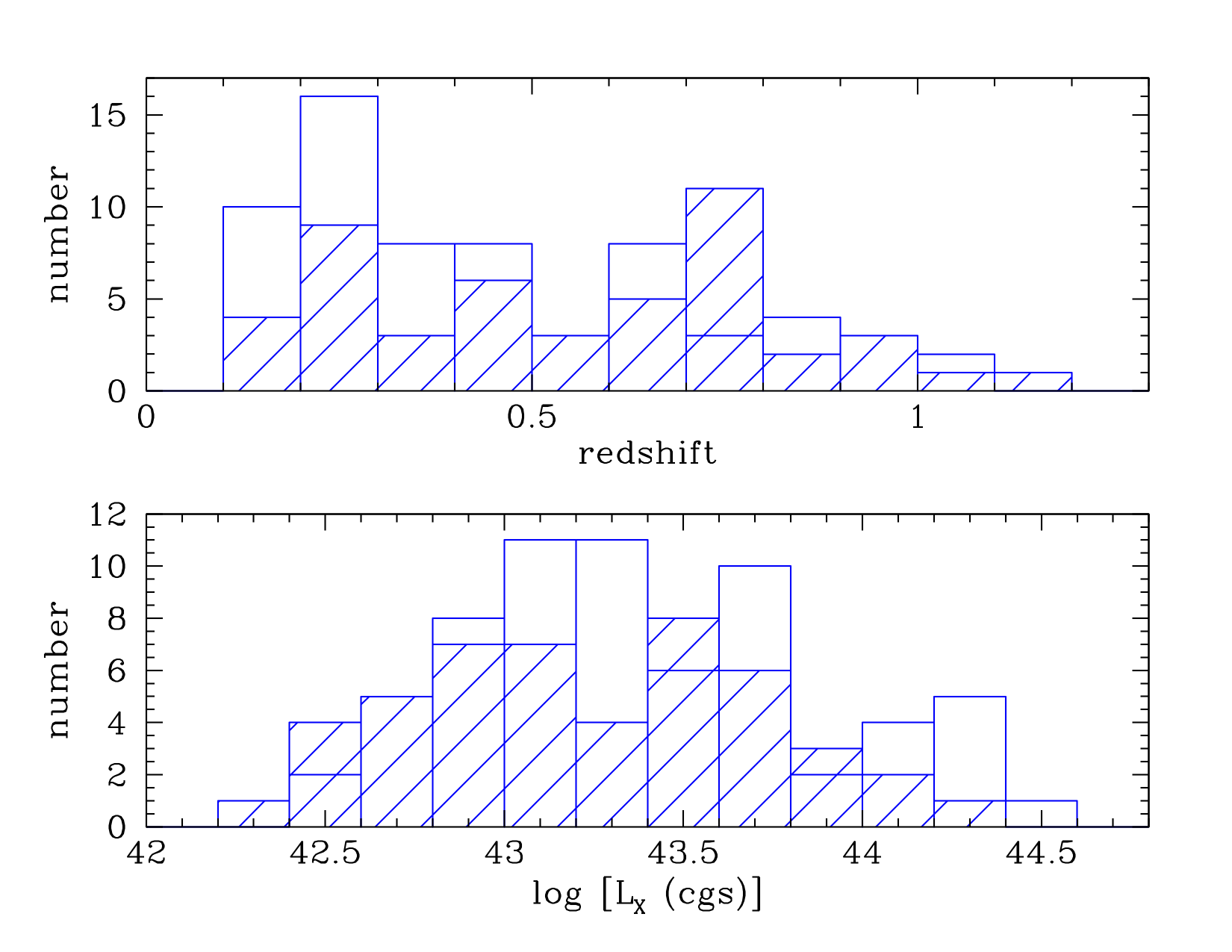}
\caption{Histograms CAMIRA high flux clusters as a function of
  redshift and X-ray luminosity. Matched and unmatched clusters are
  represented by open and shaded histograms respectively.}
\label{fig_nz_lum_by_flux}
\end{figure}

Figure \ref{fig_nz_lum_by_flux} demonstrates that the redshift and
$L_X$ distributions of the matched and unmatched high flux CAMIRA
samples are essentially identical.  A 2-sided Kolmogorov-Smirnov (KS)
test applied to the redshift and luminosity distributions respectively
generates $p$-values that the two samples are drawn from the same
population of 0.16 and 0.43.  Therefore, while it is advantageous to
present cluster sub-sample definitions in terms of intrinsic cluster
properties, e.g. X-ray luminosity, it is clear that application of a
flux threshold identifies samples of matched and unmatched CAMIRA
clusters that are comparable in terms of their X-ray luminosities.
One question to be answered therefore is whether the high flux matched
and unmatched CAMIRA clusters display any discernable differences in
their physical properties that would explain the matching results.

The situation with the XXL clusters is more straightforward to
understand.  The CAMIRA sample to which XXL is matched displays $N>15$
by definition and the XXL matching results reflect the effect of this
threshold.  There are 71 XXL clusters displaying $N>15$, of which 67
are matched to a CAMIRA cluster. The 4 unmatched clusters are either
affected by local, bright stars or are at the extremes of the CAMIRA
redshift selection interval. There are 79 XXL clusters that display
$N<15$ and none of these is matched to a CAMIRA cluster.  A second
question that we will investigate in this paper concerns the
properties of the matching clusters between XXL and CAMIRA and whether
there exist any subtle differences between them caused by the effects
of X-ray versus optical selection methods.

We therefore define the following cluster sub-samples that form the
basis for further investigation in this paper (see Table
\ref{tab_match}).
\begin{enumerate}
\item XXL $N>15$: These X-ray selected clusters exceed the CAMIRA
  catalogue richness threshold and would normally be expected to be
  detected by the CAMIRA algorithm as an optical cluster. These
  clusters are referred to as ``XXL $N>15$'' in the rest of the paper.

\item XXL $N<15$: These X-ray selected clusters do not exceed the
  CAMIRA catalogue richness threshold and would not normally be
  expected to be associated with an optically-detected cluster. These
  clusters are referred to as ``XXL $N<15$'' in the rest of the paper.

\item CAMIRA $\log f_X > -14.2$ (cgs): This flux limit contains 63/65
  CAMIRA clusters matched to an XXL cluster. These optically-selected
  clusters therefore display comparable X-ray fluxes to the XXL $N>15$
  sample and would nominally be expected to be identified as an X-ray
  cluster. Determining why 43 out of 107 CAMIRA clusters satisfying
  this flux limit are not matched to an XXL cluster is therefore of
  interest. These clusters are referred to as ``high flux CAMIRA'' in
  the rest of the paper.

\item CAMIRA $\log f_X < -14.2$ (cgs): These optically-selected
  clusters display lower X-ray flux values compared to the XXL $N>15$
  sample and would nominally not be expected to be identified as an
  X-ray cluster by the XXL pipeline. These clusters are referred to as
  ``low flux CAMIRA'' in the rest of the paper.

\end{enumerate}

\subsection{Scaling relations}
\label{sec_scaling}

We derive scaling relations between XXL X-ray aperture luminosity and
CAMIRA richness for the matched and unmatched XXL and CAMIRA samples
using a Bayesian hierarchical method with latent variables. In common
with \cite{rozo2014b} we note that the scaling relations derived in
this paper do not include any explicit information on the selection
function for either the XXL or CAMIRA surveys.  Instead, we employ the
relative scaling relations derived for the matched and unmatched XXL
and CAMIRA samples as a means of investigating whether each sample
represents a single, coherent population of objects irrespective of
whether they are matched or not.

The Bayesian fitting method employed here can deal with
heteroscedastic and possibly correlated measurement errors, intrinsic
scatters, upper and lower limits, systematic errors, missing data,
forecasting, time evolution, and selection effects. A full description
can be found in
\citet{se+et15_comalit_I,ser+al15_comalit_II,ser15_comalit_III,se+et15_comalit_IV,se+et17_comalit_V,xxl_XXXVIII_ser+al19}
(also known as \citetalias{xxl_XXXVIII_ser+al19}), which we refer to
for details. In summary, we model the relation between richness and
luminosity as a power-law with lognormal scatter. In formulae,
\begin{eqnarray}
\log_{10}(L_X/10^{42} \, \text{erg} \, \text{s}^{-1})   & = & \alpha_{L_X|Z} + \beta_{L_X|Z} Z \pm \sigma_{L_X|Z} \\
\log_{10}(N/20) & = & Z \pm \sigma_{N | Z}.
\end{eqnarray}
By the notation $\pm \sigma$, we mean that the relations are affected
by a normal intrinsic scatter with standard deviation $\sigma$. The
variable $Z$ is the latent richness, which can differ from the
observable one due to the intrinsic scatter $\sigma_{N | Z}$.
Although no explicit information on the XXL and CAMIRA selection
functions is included in this analysis, the approach by which the
independent variable $Z$ is selected from a non-evolving Gaussian
distribution provides a valid representation of the effects of a
selection threshold in the mass-observable plane \citep[see Appendix
  A1 of][]{se+et15_comalit_I}.  We consider standard priors, see
e.g. \citealt{se+et15_comalit_IV}.

For the richness, we consider a Poissonian uncertainty. For the
unmatched clusters, we consider an upper limit in the detection, see
App.~\ref{sec_uppe}. Computations were performed with the
\textsc{R}-package \texttt{LIRA}, see
App.~\ref{sec_uppe}.\footnote{The package \texttt{LIRA} (LInear
  Regression in Astronomy) is publicly available from the
  Comprehensive R Archive Network at
  \url{https://cran.r-project.org/web/packages/lira/index.html}. For
  further details, see \citet{ser16_lira}.}

\begin{figure}
\centering
\includegraphics[width=3.5in]{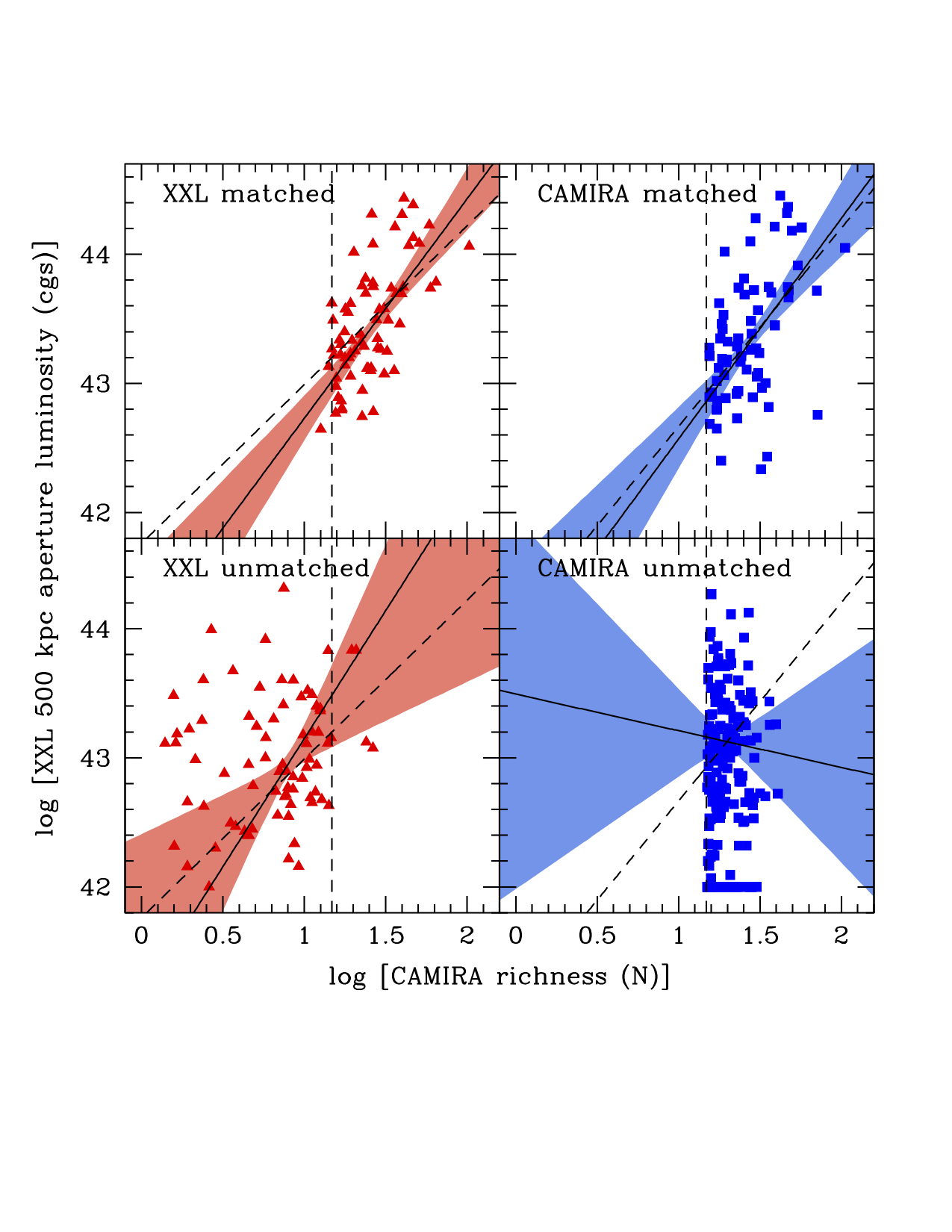}
\caption{X-ray aperture luminosity versus richness for XXL (red) and
  CAMIRA (blue) clusters. The vertical dashed line in each panel
  indicates the $N=15$ richness cut applied to generate the CAMIRA
  sample. The horizontal group of points at $\log L_X=42$ (cgs)
  represents X-ray undetected sources. The angled solid line in each
  panel represents the central scaling relation fit to each sample of
  clusters. The shaded region represent the 1-$\sigma$ confidence
  interval about the central fit. The angled dashed line in each panel
  represents the central fit to the merged (i.e. matched plus
  unmatched) sample.}
\label{fig_lum_rich_lira}
\end{figure}

Figure \ref{fig_lum_rich_lira} displays the central scaling relation
fits and their uncertainties for each sample of clusters. These
results are also detailed in Table \ref{tab_lira_results}.  It is
interesting to note that this analysis generates a slope
($\beta_{L_X|Z}$) for the relation between $L_X$ and richness for the
XXL ($1.23\pm0.12$) and CAMIRA ($1.54_{-0.24}^{+0.32}$) merged samples
that is essentially identical to that reported by \cite{rozo2014b} for
a comparison of redMaPPer and Meta-Catalogue of X-ray Clusters (MCXC;
$1.23\pm0.12$). Some caution is required however, as neither this
analyis, nor that of \cite{rozo2014b}, attempts to model any selection
effects.

A more detailed discussion of the scaling relation fits will be
presented in Section \ref{sec_discussion}. However, at this point, we
consider whether the scaling relation analysis informs the question of
whether the matched and unmatched clusters of either the XXL or CAMIRA
samples can be considered as a single population in terms of their
$L_X$-richness scaling.  The $L_X$--richness scaling relations of the
matched, unmatched and merged XXL sample are all consistent with one
another \---\ as one might expect, given that the XXL sample is X-ray
selected and presents a continuous range of richness values. When
matched to richness-selected CAMIRA clusters the matching results are
strongly correlated with richness about the $N=15$ threshold applied
to the CAMIRA catalogue.  Though the scaling relations determined for
the CAMIRA matched and unmatched samples are statistically different
(with a large scatter in particular for the unmatched clusters), it is
noteworthy that the scaling relations for the matched and merged
samples (respectively containing 68 and 289 clusters) are very close
in their values of normalisation, slope and scatter. This result would
appear to support the assertion that the CAMIRA cluster sample
represents a single, uniform population of galaxy clusters \---\ at
least as characterised on the $L_X$--richness plane. It is therefore
interesting to consider in the following sections why a large fraction
of the CAMIRA clusters do not appear to be matched to an XXL counterpart.

\begin{table*}
\caption{Scaling relation fits to XXL and CAMIRA cluster
  samples. Errors represent inter-quartile distances about the best fit (median posterior probability).}
\label{tab_lira_results}      
\centering                          
\begin{tabular}{|l|c|c|c|c|c|}        
\hline\hline                 
Sample & $N_{cluster}$ & $\alpha_{L_X|Z}$ & $\beta_{L_X|Z}$ & $\sigma_{L_X|Z}$ & $\sigma_{N|Z}$ \\
\hline
\T XXL matched & 67 & $1.24_{-0.05}^{+0.05}$ & $1.70_{-0.25}^{+0.26}$ & $0.05_{-0.02}^{+0.03}$ & $0.03_{-0.01}^{+0.02}$ \\
\T XXL unmatched & 83 & $1.74_{-0.22}^{+0.31}$ & $1.98_{-0.53}^{+0.77}$ & $0.23_{-0.10}^{+0.08}$ & $0.10_{-0.06}^{+0.06}$ \\
\T \B XXL merged & 150 & $1.36_{-0.03}^{+0.03}$ & $1.23_{-0.10}^{+0.12}$ & $0.07_{-0.03}^{+0.05}$ & $0.05_{-0.02}^{+0.03}$ \\
\hline
\T CAMIRA matched & 67 & $1.08_{-0.07}^{+0.06}$ & $1.71_{-0.33}^{+0.33}$ & $0.10_{-0.04}^{+0.08}$ & $0.05_{-0.02}^{+0.03}$ \\
\T CAMIRA unmatched & 203 & $1.12_{-0.03}^{+0.03}$ & $-0.28_{-0.71}^{+0.66}$ & $0.06_{-0.02}^{+0.04}$ & $0.04_{-0.01}^{+0.01}$ \\
\T \B CAMIRA merged & 270 & $1.13_{-0.03}^{+0.03}$ & $1.54_{-0.24}^{+0.32}$ & $0.08_{-0.03}^{+0.05}$ & $0.04_{-0.02}^{+0.03}$ \\
\hline                                   
\end{tabular}
\end{table*}

\section{Results}
\label{sec_results}

Having defined each cluster sub-sample in Section \ref{sec_subsamp} the
next task is to determine whether each sub-sample presents measureable
physical differences with respect to the others and what the cause of
these differences might be. In this section we therefore report on the
set of measurements performed on each cluster sub-sample and present
the results. We defer a discussion of these results in the context of
each cluster sub-sample until Section \ref{sec_discussion}.

\subsection{Cluster redshift distributions and visual assessment}

\begin{figure*}
\centering
\includegraphics[width=7.0in]{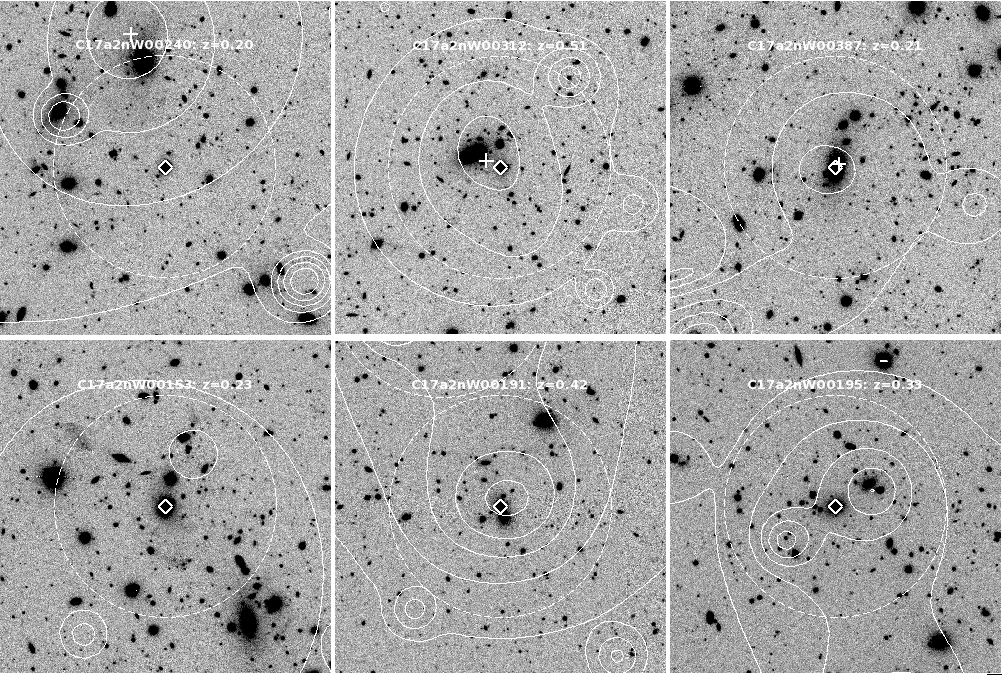}
\caption{HSC $i$-band images of examples of clusters drawn from each
  sub-sample. Each image is three arcminutes on a side with north up
  and east left. XMM emission contours are shown in white. The XMM
  data in these images are processed using version 4 of the {\tt Xamin}
  pipeline that combines individual pointings in to a mosaic. The
  dashed circle in each panel has a radius of 1 arcminute and is
  centered on the cluster location. The centroid of each CAMIRA
  detection is marked with a white diamond symbol. The centroid of
  each XXL detection is marked with a white cross symbol. Top panel:
  CAMIRA $\log f_X>-14.2$ (cgs) matched. Bottom panel: CAMIRA $\log
  f_X>-14.2$ (cgs) unmatched.}
\label{fig_vis1}
\end{figure*}

\begin{figure*}
\centering \includegraphics[width=7.0in]{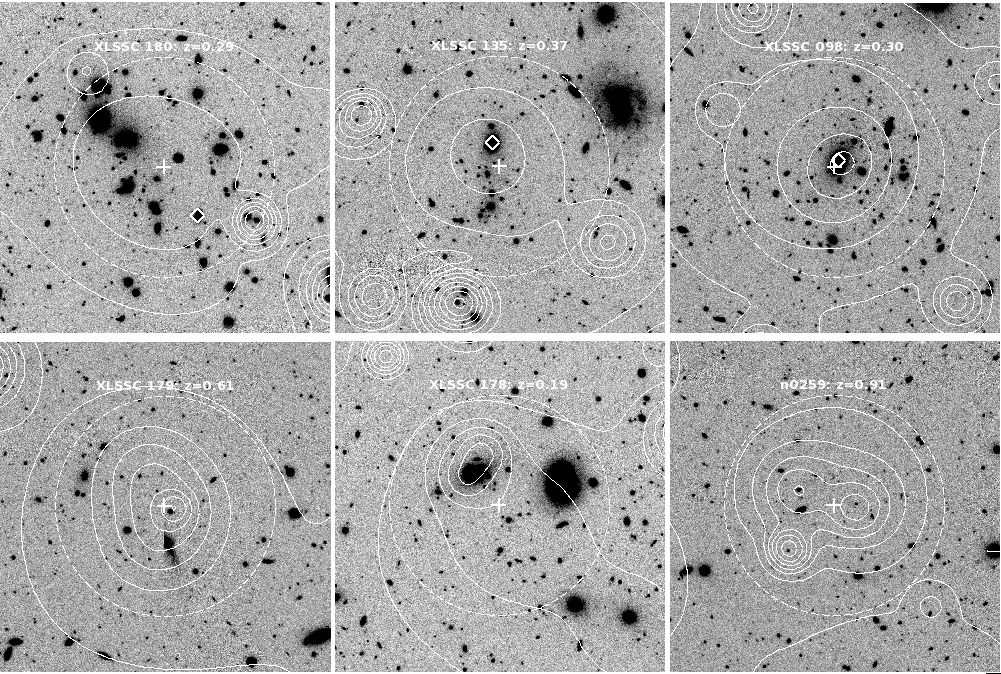} 
\caption{HSC $i$-band images of examples of clusters drawn from each
  sub-sample. Each image is three arcminutes on a side with north up
  and east left. XMM emission contours are shown in white. The XMM
  data in these images are processed using version 4 of the {\tt Xamin}
  pipeline that combines individual pointings in to a mosaic. The
  dashed circle in each panel has a radius of 1 arcminute and is
  centered on the cluster location. The centroid of each CAMIRA
  detection is marked with a white diamond symbol. The centroid of
  each XXL detection is marked with a white cross symbol. Top panel:
  XXL $N\ge15$ matched. Bottom panel: XXL $N<15$.}
\label{fig_vis2}
\end{figure*}

\begin{figure}
\centering
\includegraphics[width=3.5in]{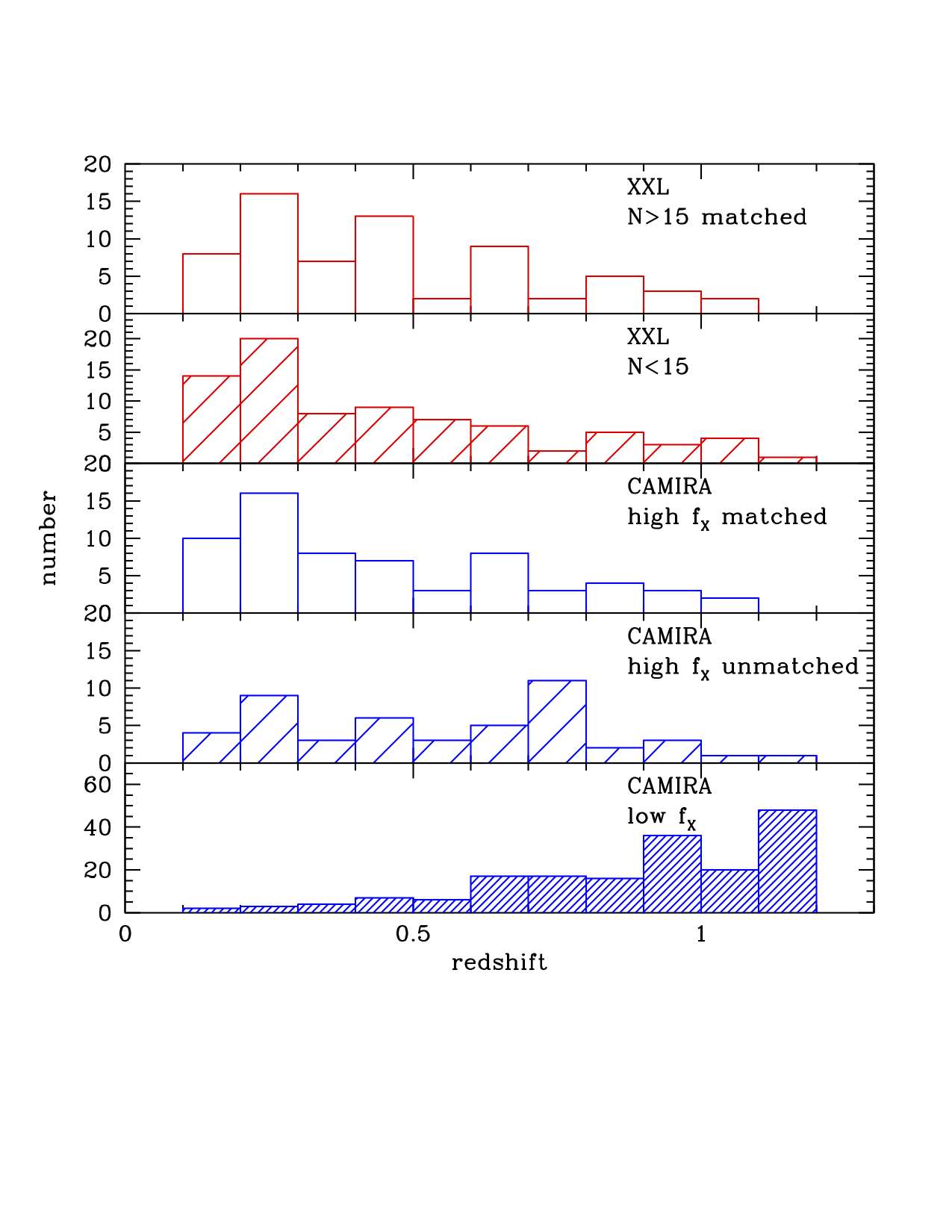}
\caption{Redshift distributions for all cluster sub-samples.  The
  redshift distributions of the XXL $N>15$ matched and CAMIRA
  high-flux matched samples are statistically identical modulo a small
  redshift scatter (see Section \ref{sec_xxl_rich}. The low-flux CAMIRA
  clusters are largely unmatched (160/163 clusters) to XXL clusters as
  they are less likely to be detected above the nominal XXL cluster
  sample flux threshold (see Section \ref{sec_subsamp}).}
\label{fig_nz_all}
\end{figure}

In Figures \ref{fig_vis1} and \ref{fig_vis2} we show example images of
clusters drawn from each sub-sample.
In addition, in Figure \ref{fig_nz_all} we plot the redshift
histograms for each cluster sub-sample.

\subsection{Stacked X-ray surface brightness profiles}
\label{sec_xsb}

\begin{figure}
\centering
\includegraphics[width=3.5in]{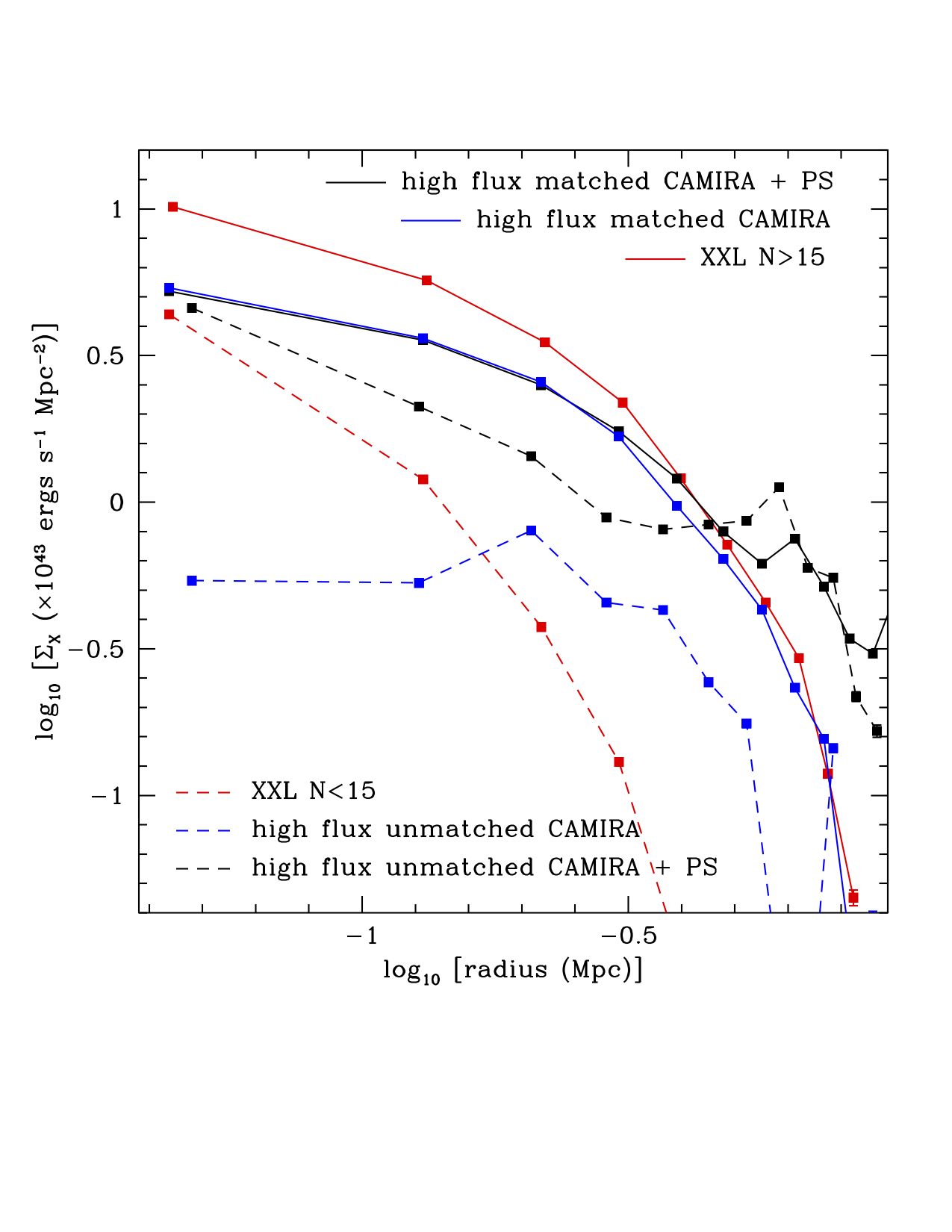}
\caption{X-ray surface brightness distributions for each cluster
  sub-sample. Profiles are displayed for XXL $N>15$ (solid red line)
  and $N<15$ (dashed red line) together with high flux CAMIRA matched
  (solid blue line) and unmatched (dashed blue line).  We additionally
  plot the surface brightness profiles for the high-flux CAMIRA
  samples having removed the point source (PS) rejection criterion
  from the stacking procedure (solid and dashed black lines for
  matched and unmatched CAMIRA clusters respectively). Note that the
  high flux CAMIRA matched sample including point sources is
  essentially identical at small radius to the same sample excluding
  point sources. Errors are not shown as they are smaller than the
  plotted symbol sizes.
}
\label{fig_xsb_all}
\end{figure}

We employ the procedure presented in \citet{willis2018} to generate a
stacked image in physical space of each cluster sub-sample defined in
Section \ref{sec_subsamp}. The stacking procedure excludes point
sources identified by the {\tt Xamin} pipeline as described in Section
\ref{sec_apphot}. However, as a test of this procedure, we also
compute stacked images for the CAMIRA cluster sub-samples without the
exclusion of point sources. We compute a circular-average surface
brightness profile for each stacked cluster sub-sample image and
present them in Figure \ref{fig_xsb_all}.

\begin{figure}
\centering
\includegraphics[width=3.5in]{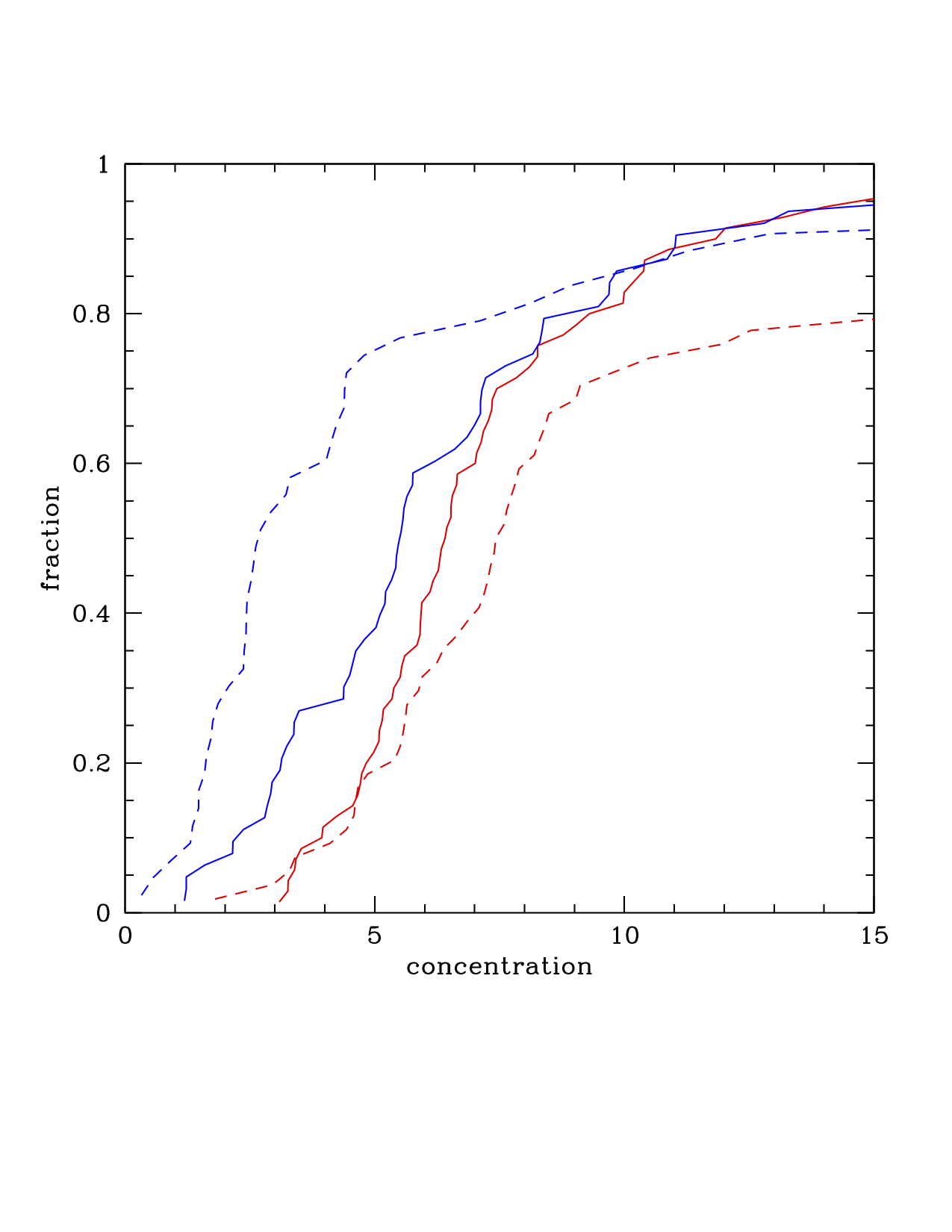}
\caption{Cumulative distribution of concentration values for each
  cluster sub-sample (see text for details). XXL $N>15$ and $N<15$ are
  displayed as solid and dashed red lines. CAMIRA high-flux matched
  and unmatched clusters are displayed as solid and dashed blue
  lines. For reference, a flat surface brightness profile will result
  in a concentration value of unity.}
\label{fig_conc}
\end{figure}

An alternative method of investigating the X-ray morphology of galaxy
clusters is to compute the concentration of X-ray emission defined as
the ratio of the X-ray surface brightness measured in two circular
apertures of differing radius \citep[e.g.][]{santos2008}. We define
concentration as the surface brightness ratio measured within circular
apertures of radius 300 and 1000 kpc centred on each cluster and
display the cumulative distribution of these values for each cluster
sub-sample in Figure \ref{fig_conc}.

\subsection{Stacked weak lensing profiles}

\begin{figure}
\centering
\includegraphics[width=3.5in]{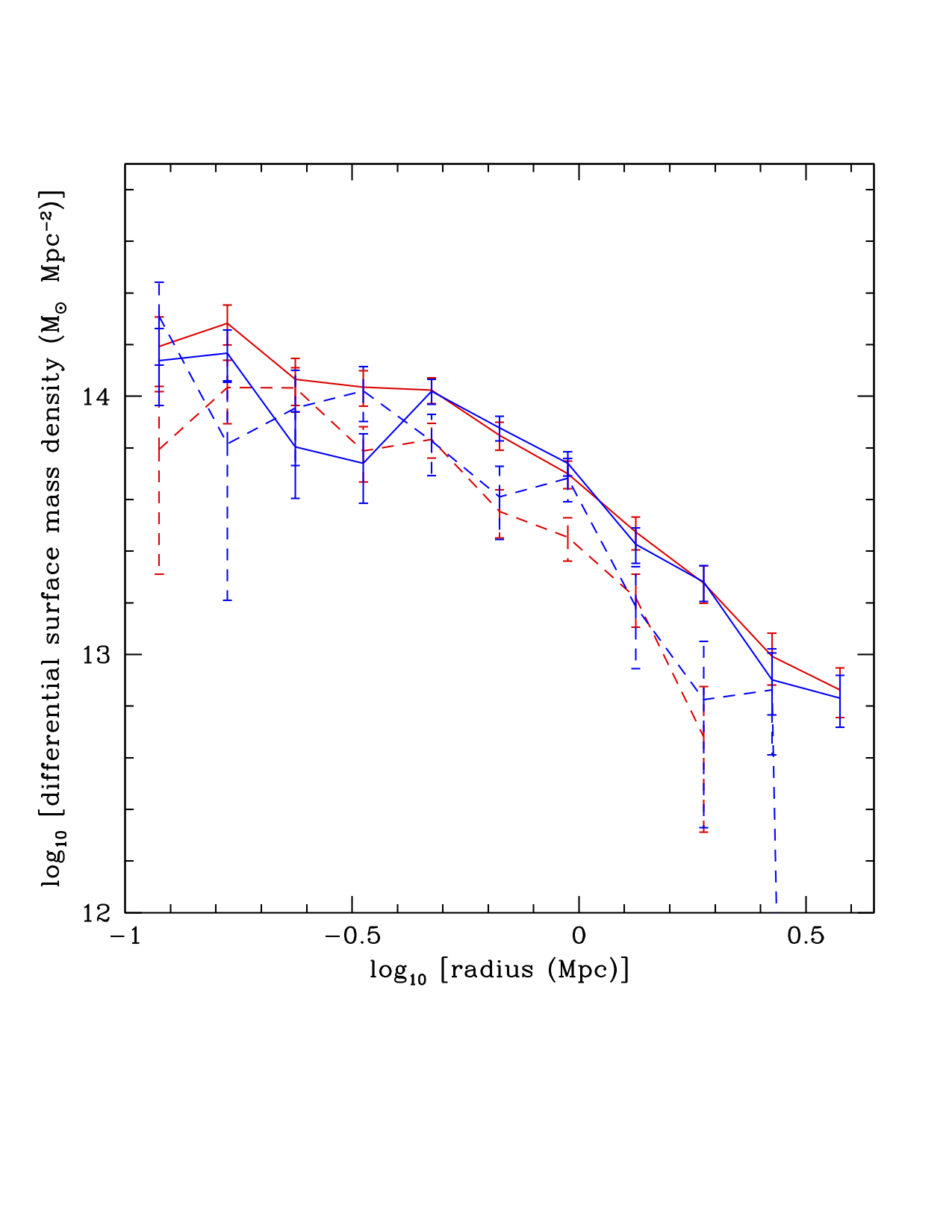}
\caption{Stacked projected weak lensing mass distributions for each
  cluster sub-sample. XXL $N>15$ and $N<15$ are displayed as solid and
  dashed red lines. CAMIRA high-flux matched and unmatched clusters
  are displayed as solid and dashed blue lines. In contrast to the
  stacked X-ray surface brightness profile (Fig. \ref{fig_xsb_all}),
  the stacked lensing profiles for all the subsample are all similar
  to each other.}
\label{fig_swl_all}
\end{figure}

We employ the HSC first-year shear catalog presented in
\citet{mandelbaum2018} to compute a stacked circular-average weak
lensing surface mass density profile for each cluster sub-sample
defined in Section \ref{sec_apphot}.  A full HSC weak-lensing analysis
of the XXL sample has been presented in \citet{umetsu2020}, which was
complemented by its companion paper, \citet{sereno2020}.  We use the
{\tt MLZ} photometric redshift \citep[see][]{tanaka2018} to estimate
the weak lensing depth, and also to remove cluster member galaxies
using the so-called P-cut method \citep{oguri2014,medezinski2018}.
Here we adopt the redshift threshold of $\Delta z=0.1$ and the
probability threshold of $p_{\rm cut}=0.95$ \citep[see][for the
  definitions of these parameters]{medezinski2018}.  Although the
choice of the parameters is less stringent than those adopted in some
of previous HSC weak lensing analysis, $\Delta z=0.2$ and $p_{\rm
  cut}=0.98$ \citep[e.g.][]{medezinski2018,miyatake2019,umetsu2020},
here we adopt this relaxed cut because we are interested in the
relative difference of mass density profiles among different cluster
subsamples rather than detailed fitting of their mass density
profiles, and because the relaxed cut helps improve the statistical
sensitivity.  Profiles are presented in Figure \ref{fig_swl_all}.

\subsection{Central galaxy offsets and member galaxy extent as measured by CAMIRA}

\begin{figure}
\centering
\includegraphics[width=3.5in]{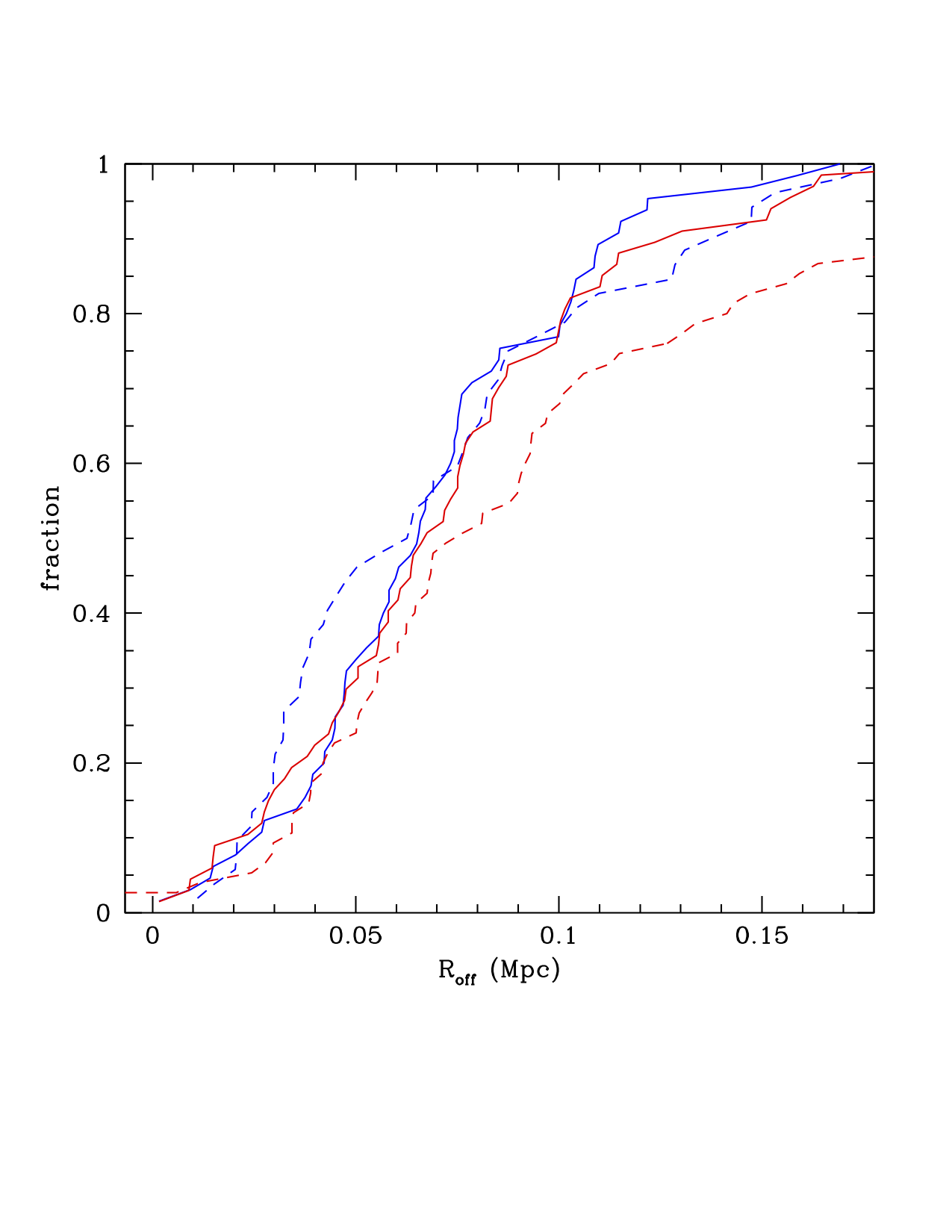}
\caption{Cumulative distributions in $R_{off}$ for XXL and CAMIRA
  clusters. XXL $N>15$ and $N<15$ are displayed as solid and dashed
  red lines. CAMIRA high-flux matched and unmatched clusters are
  displayed as solid and dashed blue lines.}
\label{fig_xxl_camira_roff}
\end{figure}

We compute the offset between the CAMIRA determined central cluster
galaxy (CCG) and the mean sky location of all cluster members
($R_{off}$). As described in Section \ref{sec_rich} the candidate CCG
in each cluster is selected as the galaxy that maximises a likelihood
function that incorporates a spatial, stellar mass and cluster
membership filter. The CCG is therefore defined as a high stellar mass
galaxy displaying a colour consistent with the cluster red-sequence
that is located close to the cluster richness peak of the richness map
\citep[see][for details]{oguri2014}.  Cumulative distributions of
$R_{off}$ for each XXL and CAMIRA cluster sub-sample are displayed in
Figure \ref{fig_xxl_camira_roff}.

\subsection{XMM off-axis angle}

\begin{figure}
\centering
\includegraphics[width=3.5in]{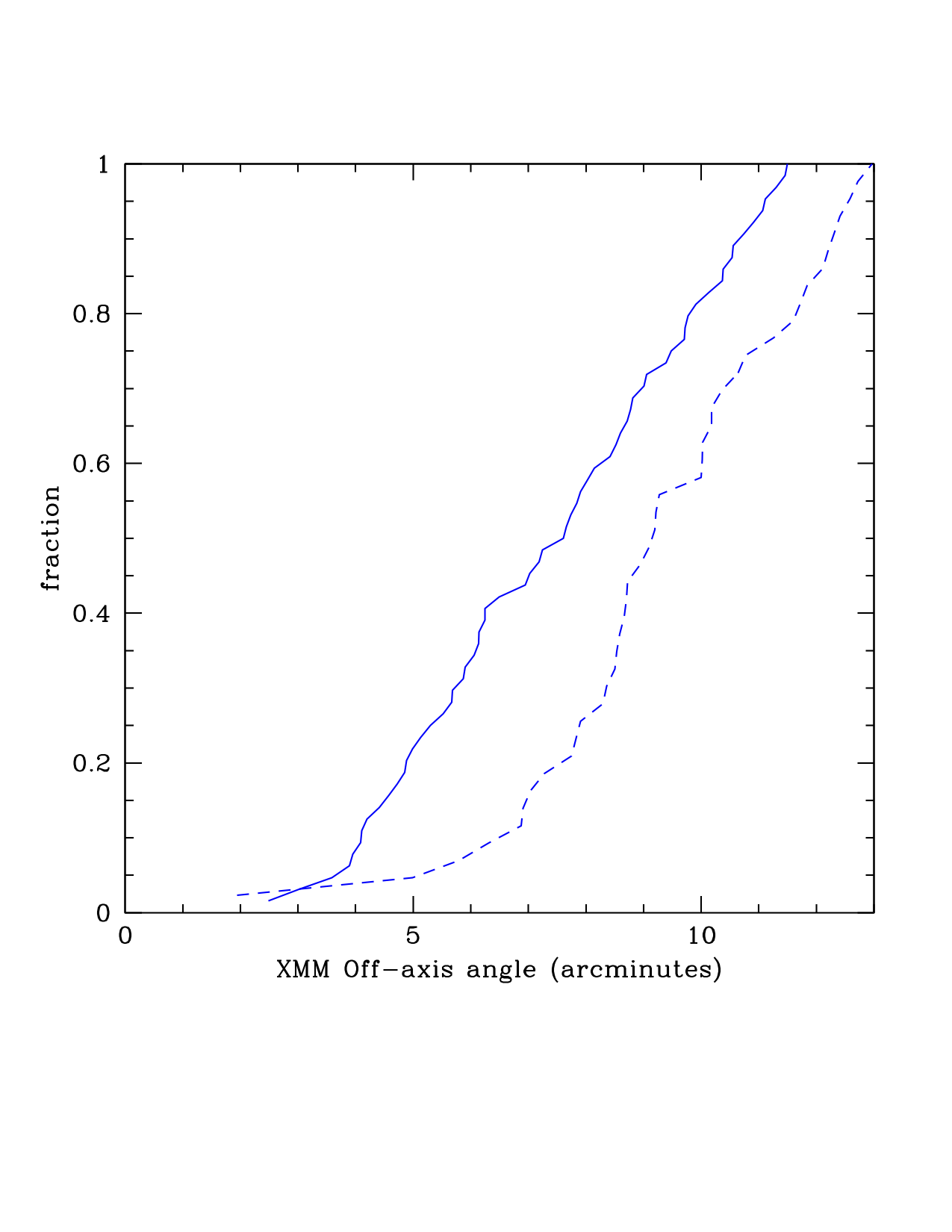}
\caption{Cumulative distributions of XMM off-axis angle for each
  CAMIRA cluster sub-sample. CAMIRA high-flux matched and unmatched
  clusters are displayed as solid and dashed blue lines.}
\label{fig_off_camira}
\end{figure}

Figure \ref{fig_off_camira} displays the cumulative XMM off-axis angle
distribution of each CAMIRA cluster sub-sample. Note that the
distribution of high-flux matched CAMIRA clusters is essentially
the same as that of the XXL $N>15$ sample (not shown).

\subsection{Point source frequency toward each cluster sub-sample}

\begin{figure}
\centering
\includegraphics[width=3.5in]{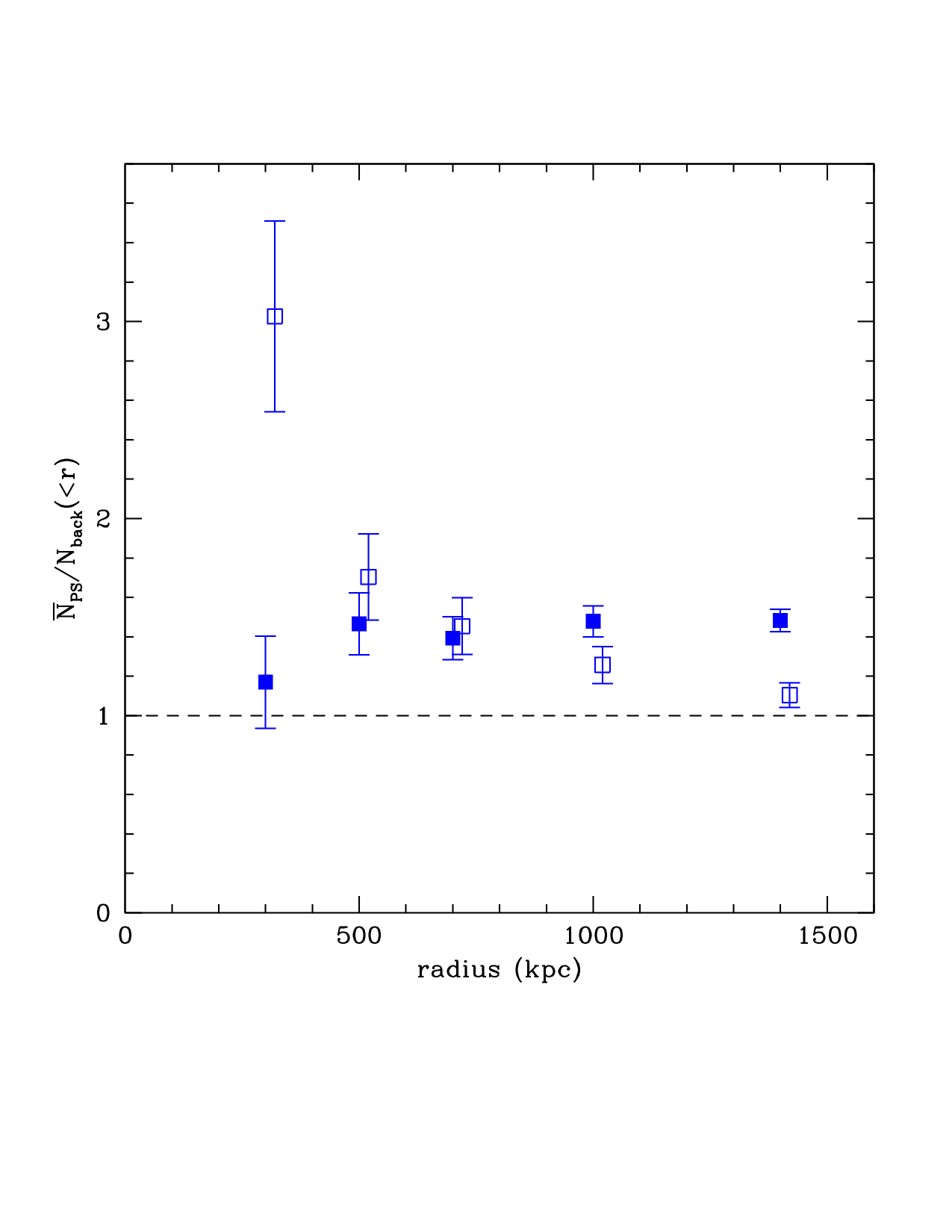}
\caption{Cumulative radial distribution of the mean number of point
  sources relative to the background for each high-flux CAMIRA
  clusters that are matched (solid squares) and unmatched (open
  squares). Values are measured at 300, 500, 700, 1000 and 1400 kpc
  for all sub-samples and points are offset in radius for
  clarity. Errors are Poissonian. For reference one would expect
  $\sim$7 point sources within a background aperture of 1400 kpc at
  the typical mean redshift of each cluster sub-sample.}
\label{fig_point_source}
\end{figure}

Figure \ref{fig_point_source} displays for each CAMIRA cluster
sub-sample the mean number of point sources (class P0 and P1) in the
3XLSS catalogue \citep[][also known as
  \citetalias{Chiappetti2018}]{Chiappetti2018} per cluster within a
given radius relative to the expectation for a background value
measured over the full XXL field.

\section{Characterising each cluster sub-sample}
\label{sec_discussion}

\subsection{XXL $N>15$ clusters}
\label{sec_xxl_rich}

\begin{figure}
\centering
\includegraphics[width=3.5in]{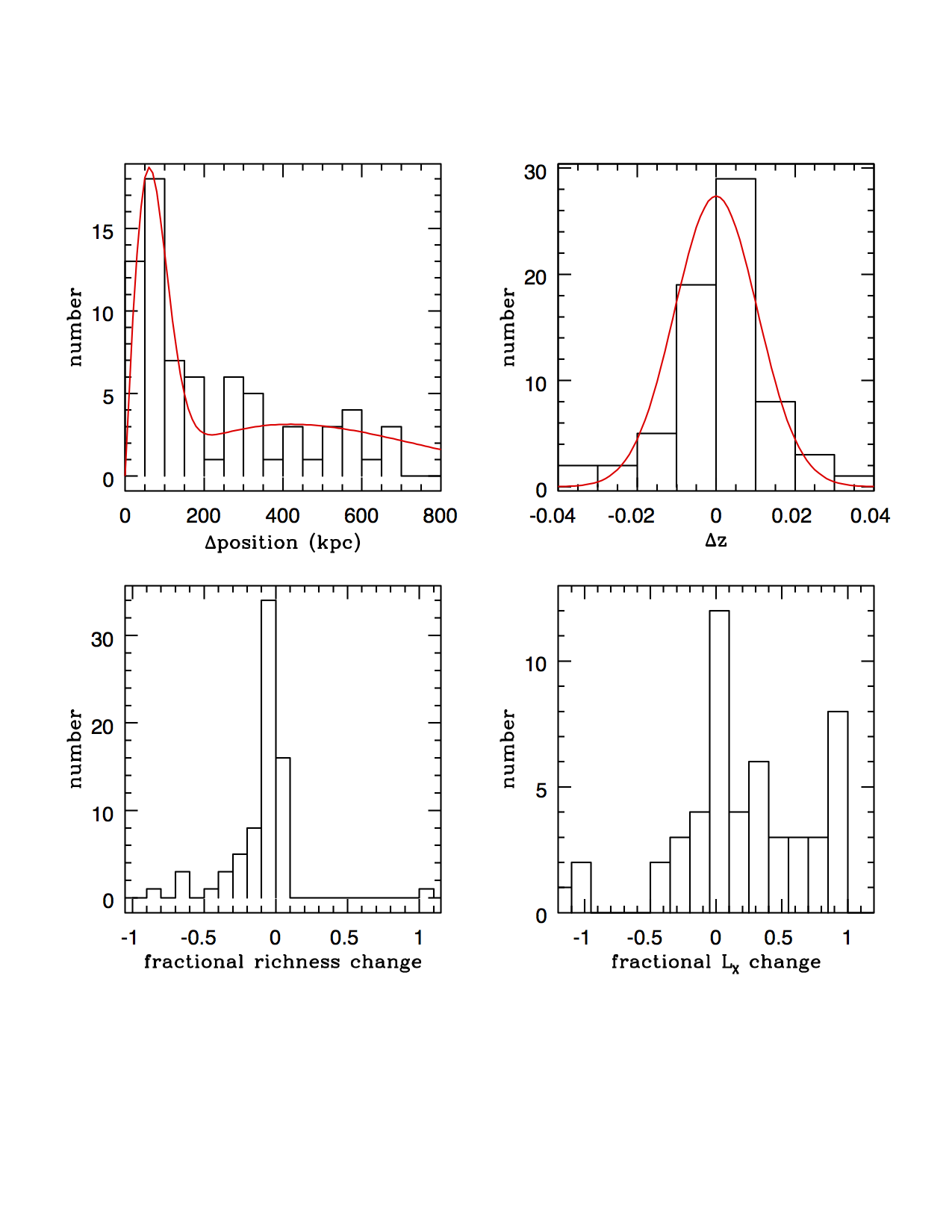}
\caption{Properties of 67 XXL $N>15$ clusters matched to CAMIRA
  selected counterparts. Top left: Histogram of rest-frame transverse
  positional offsets. The red line represents the centering model of
  \citet[][;see text for details]{oguri2014}. Top right: Histogram of redshift
  differences between matched clusters. The red line is a Gaussian
  model of zero mean and $\sigma=0.011$. Bottom left: Histogram of
  fractional richness changes between the XXL and CAMIRA-determined
  cluster locations. The convention is [XXL-CAMIRA]/XXL. Bottom right:
  Histogram of fractional change in $L_X$ measured at each cluster
  location following the same convention as above.}
\label{xxl_rich_compare}
\end{figure}

The optically rich XXL clusters are defined as those displaying
$N>15$, almost all of which are located in the top left panel of
Figure \ref{fig_flux_rich_all}. With only four exceptions they are all
matched to a CAMIRA counterpart and therefore are the same clusters as
are displayed in the top right panel of the same figure.  Figure
\ref{xxl_rich_compare} shows the properties of XXL $N>15$ clusters
matched to CAMIRA clusters within 700 kpc. The distribution of
projected transverse separations between the XXL and CAMIRA cluster
position is shown in the top left panel. Over-plotted in red is a
mis-centering model described in \citet{oguri2014} and described by
the parameters $f_{cen} = 0.45$, $r_{s,cen}= 60$ kpc and $r_s = 420$
kpc where the probability of a given centroid offset, $r$ is
\begin{equation}
{
p(r) = f_{cen} \frac{r}{r_{s,cen}} \exp \left ( - \frac{r^2}{2 r_{s,cen}^2} \right ) + (1-f_{cen}) \frac{r}{r_{s}} \exp \left ( - \frac{r^2}{2 r_{s}^2} \right ).
}
\end{equation}
The properties of this fit are different to that presented in
\citet{oguri2014} (which are based upon a comparison to XCS and ACCEPT
X-ray clusters). We measure a lower fraction of centred clusters,
$f_{cen}$ (0.45 compared to 0.7), yet mis-centered clusters are
generally observed to display the same scatter in position ($r_s =
420$ kpc).
The top right panel of Figure \ref{xxl_rich_compare} displays the
difference in redshift between the XXL values (spectroscopic) and
CAMIRA (photometric). Over plotted in red is a Gaussian model of mean
zero and standard deviation 0.011 indicating that the CAMIRA cluster
photometric redshifts appear to be very reliable.
The lower panels of Figure \ref{xxl_rich_compare} show the fractional
difference in richness and X-ray aperture luminosity between the XXL
and CAMIRA cluster positions (following the convention
[XXL-CAMIRA]/XXL). These distributions indicate that the XXL position
identifies the location of marginally greater X-ray luminosity while
the CAMIRA position traces the location of greater richness.

As noted by \cite{zhang2019} the determination of a cluster centre
based upon an optically identified dominant central galaxy is subject
to a number of uncertainties. These uncertainties are associated with
either the displacement of the central galaxy from the centre of the
cluster gravitational potential during a cluster scale merger event
\citep[e.g.][]{lavoie2016} or the mis-identification of the central
galaxy due either to the existence of multiple unmerged central
galaxies from progenitor clusters or from projection effects
\citep[e.g.][]{myles2020}.

Of the four $N>15$ XXL clusters not matched to a CAMIRA cluster, three
are potentially affected by nearby bright star halos that may affect
the HSC photometry \citep{coupon2018} and the fourth is at $z \sim 1$
and may represent a marginal CAMIRA detection. Overall, with only four
unmatched clusters, we do not attempt any further statistical
investigation of why they are unmatched.

\subsection{XXL $N<15$ clusters}

None of the XXL $N<15$ clusters are matched to a CAMIRA cluster. This
occurence results from the $N=15$ selection cut applied to generate
the CAMIRA cluster sample. Furthermore, the $L_X$-richness scaling
relation analysis presented in Section \ref{sec_scaling} indicates
that the XXL $N>15$ and $N<15$ samples are consistent with being drawn
from the same parent sample (albeit with no correction for sample
incompleteness) even given the split in the samples at $N=15$.
Furthermore, stacked weak lensing profile of the XXL $N<15$ clusters
displays a similar shape yet lower normalisation compared to the
XXL $N>15$ clusters, a result consistent with the scenario that the
$N<15$ clusters represent lower mass counterparts of the $N>15$
clusters. While it is likely that these XXL clusters would be matched
to optically poor CAMIRA clusters were the richness selection
threshold reduced, the comparison would likely be confused by the
increased rate of false positives potentially introduced into the
CAMIRA sample by doing so.

\subsection{Unmatched high-flux CAMIRA clusters}

As described in Section \ref{sec_subsamp}, 43/107 CAMIRA clusters
exceeding the $\log f_X > -14.2$ (cgs) threshold remain unmatched to
an XXL cluster while Figure \ref{fig_nz_lum_by_flux} indicates that
matched and unmatched high-flux CAMIRA clusters display equivalent
redshift and X-ray luminosity distributions. Why are such otherwise
detectable X-ray emitting galaxy clusters not identified as extended
sources by the XXL survey?

Figure \ref{fig_xsb_all} shows the stacked X-ray surface brightness
(XSB) profiles of each cluster sub-sample and the presence of
significant extended X-ray emission in the unmatched high-flux CAMIRA
clusters indicates that \---\ as a population \---\ they are real
clusters (defined as significant galaxy overdensities associated with
a hot gaseous halo).  The important point to answer here is why these
sources are not recognised as clusters (i.e. significant extended
X-ray sources) by the XXL pipeline.

The $L_X$-richness scaling relation computed for the unmatched CAMIRA
clusters (Section \ref{sec_scaling}) is poorly constrained. However,
we note that the relation for the merged matched/unmatched CAMIRA
sample (Figure \ref{fig_lum_rich_lira}) is statistically very similar
to that obtained for the matched sample (even though one is fitting 58
compared to 289 clusters). This result would appear to support the
conclusion that the CAMIRA cluster sample represents a single
population of objects.

The stacked XSB distributions in the high-flux unmatched CAMIRA
clusters are fainter in their central regions than the high flux
matched CAMIRA clusters. Note that the fainter central XSB profile in
the unmatched clusters is unlikely to be solely due to
mis-centering. Compare the XSB profiles of XXL clusters matched to
CAMIRA in Figure \ref{fig_xsb_all} (solid red line) to that of the
CAMIRA clusters matched to XXL (solid blue line). These are the same
clusters that are detected at different positions in the X-ray and
optical. Hence, the difference between these two XSB profiles is due
to mis-centering of the CAMIRA clusters compared to XXL.

The distribution of concentration measurements for these clusters
displayed in Figure \ref{fig_conc} reinforces this impression. The
effects of mis-centering are apparent in the differences between the
distributions of the matched XXL and CAMIRA clusters (the solid red
and blue lines). In contrast to this, the unmatched high-flux CAMIRA
clusters (the dashed blue line) indicates that these clusters are
significantly less concentrated than their matched counterparts. A KS
test performed upon the concentration distributions of the matched and
unmatched high-flux CAMIRA clusters indicates a $p$-value that they
are drawn from the same population of $1.5 \times 10^{-4}$.

The stacked projected weak lensing mass density for each cluster
sub-sample (Figure \ref{fig_swl_all}) indicates that (within error
fluctuations) the XXL $N>15$ clusters together with the
high-flux matched and unmatched CAMIRA clusters show the same
projected mass density profiles. The similarity of the XXL and CAMIRA
matched clusters, within the limits of mis-centering as discussed
previously, is expected. The similarity of the projected mass profile
of the high-flux unmatched CAMIRA clusters to the matched
clusters is interesting when compared to the corresponding X-ray
surface brightness profiles (Figure
\ref{fig_xsb_all}). 

Although the relative suppression of central X-ray emission might be
taken as evidence that the high-flux matched/unmatched CAMIRA clusters
represent clusters of similar mass that are experiencing different
central ICM physics \citep[e.g.][]{sanderson2009}, the unmatched X-ray
luminous CAMIRA clusters do not appear to be disturbed according to
the measures we have available to us. As shown in Figure
\ref{fig_xxl_camira_roff}, the CCG offset distribution for all cluster
sub-samples are very similar (no signficant $p$-values are obtained
between sub-samples in 2-sided KS tests).

The relative occurrence of X-ray point sources as a function of
cluster-centric distance reveals significant differences between the
high-flux matched and unmatched CAMIRA sub-samples (Figure
\ref{fig_point_source}). The point source occurence rate in the
high-flux matched CAMIRA clusters is essentially the same as the XXL
$N>15$ sample and these clusters indicate that the occurrence of point
sources in matching cluster fields is marginally elevated compared to
the level expected from the background (horizontal dashed line), yet
dips at low radius consistent with the result of
\citet{koulouridis2018} (also known as
\citetalias{koulouridis2018}). Compared to these data, the high-flux
unmatched CAMIRA clusters display a significant excess of points
sources compared to the background. The increase of this excess toward
smaller cluster-centric radii suggests that they are physically
associated with the clusters as opposed to line-of-sight projections.

The effect of point sources along the line of sight to each CAMIRA
cluster is two-fold: Firstly, the point source may simply represent
extended emission from the cluster itself that remains unclassified
due to low count rates (which would result in the source being
labelled as P0) or, more subtly, the presence of both point-like and
extended emission may result in a blended source ultimately labelled
as point-like by the pipeline (\citetalias{faccioli2018}).  Secondly,
the exclusion of point source emission from the aperture photometry
computed in Section \ref{sec_apphot} will result in the
underestimation of any extended emission within the applied aperture.

As a test of this effect we repeated the X-ray stacking procedure
described in Section \ref{sec_xsb} for the high-flux CAMIRA clusters
\---\ this time with point sources included in the stacking \---\ and
display the results in Figure \ref{fig_xsb_all}.  The XSB profile for
the high-flux matched CAMIRA clusters is largely unchanged at low
clustercentric radius and displays the effects of additional noise at
larger radius (demonstrating the motivation for originally excluding
point sources).  The XSB profile for the high-flux unmatched clusters
is significantly changed with the inclusion of point sources and
displays markedly elevated levels of X-ray emission at low
clustercentric radius.  The XSB profiles of matched and unmatched
high-flux CAMIRA clusters now appear more similar, though the
unmatched clusters are always slighlty fainter at all radii.  Although
this test confirms the effect of central point sources upon the
characterisation of high-flux unmatched CAMIRA clusters, it does not
resolve the question as to whether such sources represent true point
sources, i.e. central cluster AGN, or compact, yet extended, central
X-ray emission that remains unclassified in XMM images.

In order to resolve this question we computed the hardness ratio of
the stacked X-ray emission from each cluster sub-sample generated in
this case with no point source rejection. Following
\cite{anderson2013} we compute the X-ray hardness ratio as
\begin{equation}
  {
HR = \frac{H-S}{H+S},
  }
\end{equation}
within a circular aperture of radius 150 kpc centred on each
stack. This physical scale represents an angular scale approximately
equal to 1.5 times the on-axis Half-Energy Width (HEW) of the XMM
Newton detectors computed at a redshift $z=0.5$. We employ the [0.5-2]
keV and [2-10] keV observed frame energy intervals to represent the
count rates in the soft and hard bands respectively.

The hardness ratios computed for the high-flux matched and unmatched
CAMIRA clusters are respectively $-0.58 \pm 0.02$ and $-0.53 \pm
0.05$. For reference, canonical APEC plasma models for thermal
emission from a $T=2$ keV galaxy cluster and a simple AGN model
consisting of an absorbed power law with an index of $-2$, both
computed at $z=0.5$, generate HR values of -0.80 and -0.32
respectively. Unsurprisingly, the high-flux matched CAMIRA clusters
present a mix of hard and soft X-ray emission, hosting as they do a
mix of thermal ICM and point-like AGN emission. What is important is
that the X-ray hardness ratio of stacked emission from the high-flux
unmatched clusters is statistically identical (at the $1.5 \sigma$
level ) to that of the matched clusters. This result argues that the
high-flux unmatched CAMIRA clusters are unmatched due to the
mis-classification of extended thermal emission as point-like as
opposed to such clusters being dominated by bright, central AGN,
i.e. true point sources. However, we note that this test does not
permit us to quantify the extent to which the intermediate case
\---\ where extended emission is blended with point-like emission from
proximate AGN \---\ plays a role in the misclassification of an
extended, thermal source.

Finally, it also appears that instrumental effects also play a role in
the absence of an XXL cluster identification at these locations.
Figure \ref{fig_off_camira} indicates that high-flux unmatched CAMIRA
clusters are identified at greater XMM off-axis angle than their
matched counterparts. The KS $p$-value that the matched and unmatched
clusters are drawn from the same sample is $1 \times 10^{-3}$.  Being
located at greater off-axis angle will result in a decreased detection
probability due to a combination of vignetting and deteriorating
point-spread function (PSF). We note that, as we have not attempted to
deconvolve the effects of the PSF from the XSB distributions present
in Figure \ref{fig_xsb_all}, there exists the possibility that the
lower central X-ray surface brightness observed in the high-flux
unmatched CAMIRA clusters partly results from the larger PSF which
exists at greater XMM off-axis angle.

It therefore appears that two principal factors may act in combination
to reduce the likelihood that CAMIRA clusters are identified as
extended X-ray sources by the XXL pipeline.  Extended, thermal X-ray
emission is present in these clusters. However, when that emission is
potentially blended with proximate AGN and combined with the larger
XMM PSF at increased XMM off-axis angle, it results in a
morphologically complex source that is not recognised as extended by
the XXL pipeline.

\subsection{Low flux CAMIRA clusters}

The 176 low flux CAMIRA clusters that, with two exceptions, remain
unmatched to an XXL cluster are preferentially located at higher
redshift than all other cluster sub-samples (Figure \ref{fig_nz_all}).
While some of these clusters do indeed display X-ray flux values
comparable to some of the very faintest XXL clusters (Figure
\ref{fig_flux_rich_all}), the simplest explanation for the absence of
an XXL-detected cluster at these locations is that these clusters are
low-to-moderate X-ray luminosity sources viewed at high redshift.  As
such they present X-ray count rates that are insufficient to generate
a statistically acceptable characterisation as either extended (C1 or
C2) or point-like (P1) and are classified as P0 as a result.

\section{Conclusions}
\label{sec_conc}

The ability to effectively sample any population of objects in the
universe often reduces to a discussion of purity \---\ the ability to
distinguish true sources from false \---\ and completeness \---\ the
ability to identify as large a fraction of true sources as possible.

The XXL cluster sample represents an exceptionally pure sample of
galaxy clusters.  This statement is based upon the results of
spectroscopic follow-up of XXL galaxy clusters
(\citetalias{adami2018}), of which 95\% possess a spectroscopic
redshift.  It is therefore unsurprising that effectively all XXL
$N>15$ clusters are matched to a CAMIRA cluster.  The high
spectroscopic completeness of the XXL sample further supports the idea
that XXL $N<15$ clusters \---\ which are unmatched to a CAMIRA cluster
by virtue of the CAMIRA catalogue richness cut \---\ are real clusters
presenting low richness values consistent with the fitted
$L_X$-richness scaling relation.

In comparing CAMIRA clusters to XXL counterparts one can in principle
learn of the purity and completeness of the CAMIRA sample relative to
XXL.  A large fraction (163/270) of CAMIRA clusters \---\ which we
label as low-flux unmatched CAMIRA \---\ are simply too faint to be
characterised as extended by the XXL pipeline.  Flux incompleteness is
a well-studied selection effect and is modelled explicitly in the XXL
pipeline (\citealt{pacaud2006}; \citetalias{pacaud2016new};
\citetalias{faccioli2018}).

However, we find that a further 40\% (43/107) of high flux CAMIRA
clusters are not matched to an XXL cluster. These CAMIRA clusters are
likely real in that each represents a galaxy overdensity associated
with significant extended X-ray emission and weak lensing mass.  The
X-ray flux threshold applied in this paper to understand such clusters
identifies 96\% (64/67) of CAMIRA clusters that are matched to an XXL
cluster. To understand why a large fraction of the remaining high flux
CAMIRA clusters are not classified as a C1/C2 source within XXL one
must recall that, to achieve high purity, the XXL pipeline selects
only bright, significantly extended sources
(e.g. \citealt{pacaud2006}; \citetalias{pacaud2016new}).

The high-flux, unmatched CAMIRA clusters display an apparent excess of
central X-ray point sources compared to both high-flux, matched CAMIRA
clusters and the field. However, it further appears that the hardness
ratio of stacked X-ray emission from these high-flux unmatched CAMIRA
clusters is statistically identical to that measured for the high-flux
matched CAMIRA clusters (which by definition are the same as the
matched XXL clusters). There is no evidence for an excess of hard
X-ray emission in the unmatched clusters that might be expected if the
excess point sources associated with these clusters were solely due to
AGN emission. Instead it appears that the point sources in these
clusters represent extended emission that is either unclassified due
to low count rates or classified as a point source due to
blending. Due to the averaging process involved in our stacking
procedures we cannot rule out that some of these clusters contain real
point sources in addition to compact extended emission and we note
that the presence of a point source close to an extended source
further complicates the extension classification with XXL
(\citetalias{faccioli2018}). A final point to note is that the
unmatched, high-flux CAMIRA clusters lie preferentially towards the
periphery of the XMM field of view such that vignetting and a
broadened PSF reduce the probability that a compact yet extended X-ray
source will be successully classified.  Overall, there is no evidence
on the basis of the comparison in this paper that the high-flux,
unmatched CAMIRA clusters are anything but galaxy clusters that, as a
result of a combination of known selection effects, are not recognised
as extended sources in the XXL pipeline.

Issues of selection are a particular concern for studies that use
galaxy cluster surveys to infer accurately the cosmological parameters
that define our Universe \citep[see][for a review]{allen2011}.
Incomplete knowledge of the selection process will potentially result
in biased inference, e.g., if the survey selection function fails to
account for clusters underrepresented due to astrophysical and
instrumental effects, inferred parameters such as $\Omega_M$ will be
biased low \citep[e.g.][]{schellenberger2017,xu2018}.  The results of
this paper indicate that there is an important requirement to describe
accurately the classification of extended X-ray sources and proximate
X-ray point sources in simulated XMM images. Presently, X-ray point
sources are included in selection function modelling via a spatially
uncorrelated background (\citealt{clerc2014};
\citetalias{pacaud2016new}) and, though X-ray point sources can be
superimposed upon extended cluster emission
(\citetalias{faccioli2018}), these studies do not include information
on the population statistics of AGN in clusters \citep[][also known as
  \citetalias{koulouridis2018b}]{koulouridis2018b}.  The incorporation
of such information into the planned version 4 processing of the XXL
survey, in addition to updates to classify sources using mosaiced
tiles of XMM images as opposed to individual pointings, will therefore
provide an important advance in the ability of XMM-based cluster
surveys to accurately represent cluster population statistics.

It is more difficult to compare the relative purity of the two
samples. The X-ray faint CAMIRA clusters present X-ray emission
(albeit faint) and are plausibly unmatched to XXL sources simply as a
result of a combination of possessing low- to moderate X-ray
luminosity and being located at high redshift.  Therefore, while there
is some certainty that the CAMIRA cluster sample identifies a greater
fraction of clusters of a given X-ray luminosity than the XXL sample,
these differences lie within the realm of known selection effects. On
the other hand, the relative purity of the CAMIRA sample with respect
to XXL has not been addressed conclusively by this analysis and
remains a question better suited to analysis either via deeper X-ray
observations or realistic mock
observations \citep[e.g.][]{adam2019}.

\section*{Acknowledgements}

Based on observations obtained with XMM-Newton, an ESA science mission
with instruments and contributions directly funded by ESA Member
States and NASA. J.P.W. acknowledges support from the National Science
and Engineering Research Council of Canada. M.S. acknowledges
financial contribution from contract ASI-INAF n.2017-14-H.0 and INAF
`Call per interventi aggiuntivi a sostegno della ricerca di main
stream di INAF'. This work was supported in part by World Premier
International Research Center Initiative (WPI Initiative), MEXT,
Japan, and JSPS KAKENHI Grant Number JP18K03693. MP acknowledges
long-term support from the Centre National d'Etudes Spatiales
(CNES). The French collaborators were supported by the Programme
National Cosmology et Galaxies (PNCG) of CNRS/INSU with INP and IN2P3,
co-funded by CEA and CNES. S.E acknowledges financial contribution
from the contracts ASI-INAF Athena 2019-27-HH.0, ``Attivit\`a di
Studio per la comunit\`a scientifica di Astrofisica delle Alte Energie
e Fisica Astroparticellare'' (Accordo Attuativo ASI-INAF
n. 2017-14-H.0), INAF mainstream project 1.05.01.86.10, and from the
European Union’s Horizon 2020 Programme under the AHEAD2020 project
(grant agreement n. 871158). S.A. acknowledges support from the
Scientific and Technological Research Council of Turkey (TUBITAK) with
the project number 117F311. K.U. acknowledges support from the
Ministry of Science and Technology of Taiwan (grants MOST
106-2628-M-001-003-MY3 and MOST 109-2112-M-001-018-MY3) and by
Academia Sinica (grant AS-IA-107-M01).

XXL is an international project based around an XMM Very Large
Programme surveying two 25 deg$^2$ extragalactic fields at a depth of
$\sim 6 \times 10^{-15} \, \rm erg \, cm^{-2} s^{-1}$ in the [0.5-2]
keV band for point-like sources. The XXL website is
http://irfu.cea.fr/xxl. Multi-band information and spectroscopic
follow-up of the X-ray sources are obtained through a number of survey
programmes, summarised at http://xxlmultiwave.pbworks.com/. The Hyper
Suprime-Cam (HSC) collaboration includes the astronomical communities
of Japan and Taiwan, and Princeton University.  The HSC
instrumentation and software were developed by the National
Astronomical Observatory of Japan (NAOJ), the Kavli Institute for the
Physics and Mathematics of the Universe (Kavli IPMU), the University
of Tokyo, the High Energy Accelerator Research Organization (KEK), the
Academia Sinica Institute for Astronomy and Astrophysics in Taiwan
(ASIAA), and Princeton University.  Funding was contributed by the
FIRST program from the Japanese Cabinet Office, the Ministry of
Education, Culture, Sports, Science and Technology (MEXT), the Japan
Society for the Promotion of Science (JSPS), Japan Science and
Technology Agency (JST), the Toray Science Foundation, NAOJ, Kavli
IPMU, KEK, ASIAA, and Princeton University. This paper makes use of
software developed for the Large Synoptic Survey Telescope. We thank
the LSST Project for making their code available as free software at
http://dm.lsst.org This paper is based in part on data collected at
the Subaru Telescope and retrieved from the HSC data archive system,
which is operated by Subaru Telescope and Astronomy Data Center (ADC)
at NAOJ. Data analysis was in part carried out with the cooperation of
Center for Computational Astrophysics (CfCA), NAOJ.

The Pan-STARRS1 Surveys (PS1) and the PS1 public science archive have
been made possible through contributions by the Institute for
Astronomy, the University of Hawaii, the Pan-STARRS Project Office,
the Max Planck Society and its participating institutes, the Max
Planck Institute for Astronomy, Heidelberg, and the Max Planck
Institute for Extraterrestrial Physics, Garching, The Johns Hopkins
University, Durham University, the University of Edinburgh, the
Queen’s University Belfast, the Harvard-Smithsonian Center for
Astrophysics, the Las Cumbres Observatory Global Telescope Network
Incorporated, the National Central University of Taiwan, the Space
Telescope Science Institute, the National Aeronautics and Space
Administration under grant No. NNX08AR22G issued through the Planetary
Science Division of the NASA Science Mission Directorate, the National
Science Foundation grant No. AST-1238877, the University of Maryland,
Eotvos Lorand University (ELTE), the Los Alamos National Laboratory,
and the Gordon and Betty Moore Foundation.

\section*{Data Availability}

All XMM public data are available through the XMM archive located at
{\tt https://xmm-tools.cosmos.esa.int}. All HSC-SSP data are publicly
available at {\tt https://hsc-release.mtk.nao.ac.jp/}.



\bibliographystyle{mnras}
\bibliography{references2} 



\appendix

\section{Upper limits}
\label{sec_uppe}

In the Bayesian framework, we associate a variable $x$ to the result
of the measurement process and a variable $X$ to the ideal result we
would get in an experiment with unlimited precision. Observational
results expressed as upper limits can be dealt by truncating the
conditional probability distribution of $x$ given $X$,
\begin{equation}
\label{eq_bug_multi_2}
P(x  | X  )  \propto  {\cal N} \left( X, \delta_x \right)  \times {\cal H}(x_{\text{ul}}- x ) \ ,
\end{equation}
where $\ {\cal N}$ is the Gaussian distribution, ${\cal H}$ is the
Heaviside function, $\delta_x$ is the observational uncertainty, and
$x_\text{ul}$ is the upper limit for the left-censored point. If the
upper limit is expressed as the probability that $x$ is less than a
threshold, or if the upper limit itself is affected by some
statistical uncertainty, the truncation can be smooth
\begin{equation}  
x_\text{ul} \sim  {\cal N}  \left( X_\text{ul}, \delta_{x_\text{ul} }\right) \ ,
\end{equation}
where $\delta_{x_\text{ul} }$ sets the transition length. If unknown,
the variable $x$ can be dealt as parameters to be fitted.

The previous treartments is implemented in the \texttt{LIRA}
package. Let \texttt{x} and \texttt{y}, \texttt{delta.x} and
\texttt{delta.y}, and \texttt{y.upperlimit} be the vectors storing the
values of the observed $\boldmath{x}$ and $\boldmath{y}$, their
uncertainties $\boldmath{\delta_x}$ and $\boldmath{\delta_y}$, and the
estimated upper limits, respectively. if unknown, the $x$ or $y$
values can be stored as \texttt{NA}. For detected objects, the upper
limits can be set to \texttt{NA} or very large values. The LIRA
command to be used to reproduce our results is

\

\noindent \texttt{> mcmc <- lira (x, y, delta.x = delta.x, delta.y =
  delta.y, y.upperlimit=y.upperlimit, sigmaXI.Z.0='prec.dgamma',
  n.chains = 4, n.adapt = 4*10\^{}3, n.iter = 4*10\^{}4) }

\

\noindent where the argument \texttt{sigma.XIZ.0 = $'$prec.dgamma$'$}
makes the scatter in $X$ a parameter to be fitted with a prior on the
precision described by a Gamma distribution, and where each of the
\texttt{n.chains = 4} chain was \texttt{n.iter = 5$\times$10$^4$}
long, and the number of iterations for inizialisation was set to
\texttt{n.adapt =4*10$^3$}.


\bsp	
\label{lastpage}
\end{document}